\documentclass[lettersize,journal]{IEEEtran}
\usepackage{cite}
\usepackage{amsmath,amssymb,amsfonts,amsthm,mathtools}
\usepackage{graphicx}
\usepackage{textcomp}
 \usepackage{enumitem}
\usepackage{xcolor}
\usepackage{blindtext}
\usepackage{lipsum}
\usepackage{subfig}
\usepackage{cuted}
\usepackage{physics}
\usepackage[long]{optidef}
\usepackage[detect-none]{siunitx}
\usepackage{multirow}
\usepackage{array}
\usepackage{steinmetz}
\usepackage{bigints}
\newtheorem{theorem}{Theorem}
\newtheorem{lemma}[theorem]{Lemma}
\usepackage{epstopdf}
\usepackage{algorithm}
\usepackage{algpseudocode}
\usepackage{caption}
\usepackage{subcaption}
\usepackage{soul}

\usepackage[labelfont=normalfont, labelsep=period, subrefformat=parens, labelformat=parens]{subcaption}

\begin{document}
\title{Auction-based Adaptive Resource Allocation Optimization in Dense and Heterogeneous IoT Networks}

\author{Nirmal~D.~Wickramasinghe,~\IEEEmembership{Student~Member,~IEEE,}
John~Dooley,~\IEEEmembership{Member,~IEEE,}
Dirk~Pesch,~\IEEEmembership{Senior Member,~IEEE,} 
and Indrakshi~Dey,~\IEEEmembership{Senior Member,~IEEE}

\vspace{-5mm}

\thanks{N.~D.~Wickramasinghe is with the Department of Electronic Engineering, Maynooth University, Ireland. (Email: nirmal.wickramasinghe.2023@mumail.ie)}
\thanks{J.~Dooley is with the Department of Electronic Engineering, Maynooth University, Ireland. (Email: john.dooley@mu.ie)}
\thanks{D.~Pesch is with the School of Computer Science and Information Technology, University College Cork, Ireland. (Email: d.pesch@cs.ucc.ie)}
\thanks{I.~Dey is with the Walton Institute, South East Technological University, Waterford, Ireland. (Email: indrakshi.dey@waltoninstitute.ie)}
\thanks{This work was supported in part by Taighde Éireann - Research Ireland under Grant 13/RC/2077\_P2, by the EU MSCA Project ``COALESCE'' under Grant Number 101130739.}
}

\maketitle

\begin{abstract}
Efficient and reliable resource allocation within densely-deployed massive IoT networks remains a key challenge due to resource constraints among low size, weight and power (SWaP) IoT devices and within the network and limitations of conventional centralized methods under incomplete information. We propose a novel auction-based framework for adaptive resource allocation, combining space-time-frequency spreading (STFS) techniques with Bayesian Game approaches. We introduce novel modified Simultaneous Ascending Auction (mSAA) mechanism tailored to densely-deployed and low-complexity IoT networks, enabling distributed computation and reduced power consumption. By incorporating Bayesian game-based bidding strategies and optimizing dispersion matrices for signal transmission, the proposed approach ensures enhanced channel throughput and energy efficiency. Comparative analysis against traditional auction types, including First-Price and Second-Price Sealed-Bid Auctions, as well as the Vickrey–Clarke–Groves (VCG) mechanism, demonstrates the superiority of mSAA in terms of surplus maximization, revenue efficiency, and robustness in risk-prone bidding environments. Simulation results validate the model's adaptability to heterogeneous IoT nodes and its potential for dense deployment across different environments and verticals. 
\end{abstract}

\begin{IEEEkeywords}
\hspace{-3mm} IoT networks, Auction Game Theory, Resource Allocation, Space-Time-Frequency Spreading
\end{IEEEkeywords}
\IEEEpeerreviewmaketitle

\section{Introduction}
%
%
%
%
\IEEEPARstart{T}{he} proliferation of Internet of Things (IoT) devices is generating unprecedented volumes of data, demanding efficient resource management strategies, especially at the network's edge. Edge computing, which brings computation closer to data sources, offers the potential to reduce latency and alleviate network congestion. However, the resource-constrained nature of edge devices presents significant challenges. Optimizing resource allocation, including power, space, and time-frequency resources, is crucial for ensuring deterministic connectivity, efficient networked computing, and intelligent control in these distributed IoT networks  \cite{IoE}. 
Several studies in the literature \cite{suggest_paper_01, suggest_paper_02, suggest_paper_03} have addressed resource allocation optimization in wireless communication networks, employing both classical optimization techniques and machine learning (ML)-based approaches. In addition, dense IoT networks extensively demand massive computational complexity for the centralized resource allocation architectures, and difficulties with fair allocation tolerating for asymmetric users or environmental parameters. To address the above challenges, this paper addresses the critical theme of joint resource allocation optimization within IoT networks, particularly focusing on power optimization of individual IoT devices through space and time-frequency spreading techniques in combination with auctioning for sub-optimal allocation of resources.

\subsection{Bayesian Theory for Edge}
The increasing volumes of data generated by IoT devices necessitate efficient resource management at the network's edge. While edge computing offers solutions to latency and congestion, the resource-constrained nature of edge devices introduces complexities. Although existing literature proposes novel approaches for edge computing \cite{J_IoT_EC_QoS, J_IoT_EC_task_allocation, J_IoT_EC_survey}, it lacks discussion of transformative technologies that could harmonize diverse layers for compatibility and synergy. At each edge layer, specific competitive demands exist, and individual self-interest can negatively impact overall performance. Consequently, a core task of edge computing involves addressing sub-optimization problems driven by user interactions within competitive edge environments. In few of the recent works, modern edge computing is framed as a cooperative game model, posing challenges in discovering unique states, particularly when optimizing multi-agent systems under complex network specifications \cite{GameTheory_vs_optimization, J_IoT_GT_QoS, J_IoT_GT_pricing_RA_offloading, J_IoT_GT_Stackelberg_GT}. Therefore, the current state-of-the-art identifies this challenge of optimized allocation within a resource-constrained environment as a cooperative game model. However, such a game model poses challenges and limitations in exploring unique states while consolidating all entities, especially under the complex influences of network specifications.

The competition for resources often occurs in scenarios where individual nodes within a dense IoT network have limited awareness or visibility of their surroundings, deviating from a perfect information game model. A perfect information game assumes that all players are fully aware of all previous moves and the current situation, allowing for optimal decision-making based on complete knowledge. However, in edge computing scenarios, this assumption often does not hold, as nodes may have limited information about network conditions or the actions of other devices. Such a scenario is therefore constrained by common data distributions. To overcome these limitations, Bayesian strategies offer a way to govern sub-optimization problems by considering expected utility within a cooperative game model. Therefore, applying Bayesian game theory to edge computing protocols is desirable \cite{BGT_contract_theory_EC}. Existing work presents an incomplete game model for efficient power consumption with incomplete channel state information (CSI) in IoT uplinks \cite{our_GLOBECOM_paper}. Yet, the vast number of random network entities and complex action vectors create computational overhead for current solvers \cite{Game_Auction_th_online_dual_avg}. Hence, instead of centralized computation, employing mechanism design based on stochastic structures is more effective, avoiding game-theoretic limitations \cite{Game_Auction_theory}. Revelation principles are also key for resource allocation in the edge network, offering low computational complexity and distributed processing capabilities by extracting highly correlated network features.

\subsection{Auction Theory for Wireless Communication}
Auction theory is expected to be a vital platform for advancements in $6G$ wireless networks delivering various decision-making architectures. Authors in \cite{Auction_decision_making_01_RA} utilize auction theory, integrating with Blockchain for resource allocation in heterogeneous networks. Conventional Vickery Clarke's auction procedure 
is exploited to design a decision-making platform nourishing the trust among users and eliminating the possibility of manipulation or tampering. Hence, fairness index would be improved by acknowledging the dynamic complexity of wireless networks concerning user satisfaction level and data rates. In addition, an enhanced licensed shared access framework is implemented for mobile network operators to address the spectrum scarcity in unmanned aerial vehicular (UAV)-based networks, \cite{Auction_decision_making_02_UAV}. Moreover, \cite{Auction_decision_making_04_Secure_IoV} explains a learning-based auction scheme for reliable and secure 6G-enabled Internet of Vehicles (IoV), addressing challenges related to untrustworthy model transfer and data sharing by malicious vehicles. To mitigate non-standard user demands in the network, penalty-based resource allocation mechanisms \cite{Auction_decision_making_06_penalty} can be introduced within auction-theoretic platforms. In general, auction theory serves as an economically inspired foundation for enhancing the quality of experience (QoE) in service delivery, thereby fairly satisfying the most critical demands of all users in the 6G era, \cite{Auction_decision_making_05_user_centric_RA}.

Auction theory, employing Bayesian Games for multi-agent optimization under imperfect information, can provide a foundation for mechanism design in resource allocation protocols within dense IoT networks. Bayesian Games extend game-theoretic analysis to scenarios of incomplete information, a departure from perfect information games where all players are fully informed. It also diverges from cooperative games that assume binding agreements and joint strategies, focusing instead on individual decision-making under uncertainty. Auctions, in Bayesian Games, play a key role in price formation models, particularly in procurement, patent licensing, and public finance \cite{Auc_theory_economic_1}, \cite{Auc_theory_economic_2}. Advanced auction techniques are used in automated mechanism design, such as advertisement auctions in e-commerce platforms \cite{Auc_theory_Ad_auc}, and in real-time bidding (RTB) for e-advertising \cite{Auc_theory_ad_RTB}. Auction-based frameworks are also suitable for edge computing \cite{J_IoT_Auction_EC_FoG, J_IoT_Auction_MEC}, as seen in auto-bidding mechanisms \cite{Auc_theory_auto_bidding}. In addition, auction-driven methods address efficient resource allocation in cooperative wireless communications \cite{Auc_theory_cooperative_commn}, vehicular cloud-assisted networks \cite{Auc_theory_vehicular_cloud_assisted}, and blockchain-driven fog environments \cite{Auc_theory_blockchain_fog_computing}.

\subsection{Auction Theory for Edge}
Protocols and strategies for automated multi-attribute auctions offer a versatile foundation for advanced decision support systems. These systems address complex utility functions, including the inter-dependencies between conventional wireless network attributes \cite{Auc_theory_multi_attribute_Israel, Auc_theory_multi_attribute_ad}. The authors in \cite{Auc_theory_cloud_fog} provides a comprehensive survey on auction-based mechanisms in cloud/edge computing, focusing on efficient resource management and pricing challenges. \cite{Auc_theory_fog_cloud_offloading} highlights auctions as effective tools for incentivizing task offloading and resource allocation in fog-cloud systems, prompting the reconsideration of auction-based strategies for the edge. Auctions enable dynamic resource allocation by leveraging user synergy across sub fog layers to enhance edge performance. Each auction participant contributes to decision-making by distributing computational complexity via node collaboration. To optimize resource allocation with incomplete information in dense IoT networks, sealed-bid auction mechanisms are proposed, addressing IoT device limitations and promoting adaptive, sustainable models for the edge layer. 

    There are several sealed-bid auctions in the literature, and we can identify them via two major categories. One is the legacy of auction theory or fundamentals in sealed-bid auctions, such as the first price sealed bid (FPSB), which is generalized from the English auction and the second price sealed bid (SPSB). These two auctions are targeting single-objective auctions, and the Vickery-Clarke-Grove mechanism, which is for multiple-objective auction, is derived using SPSB as the kernel. For the second category, novel types of auctions such as combinatorial auctions (CA), double auction (DA), combinatorial double auction (CDA), and simultaneous ascending auction (SAA) have different configurations in the mechanism and are generalized via the previous fundamental auction types. To strengthen the explanation with recent literature works, we are following the latest survey (published on 15th August 2025) about Auction theory and Game theory for the edge computing, including IoT networks, \cite{Auction_survey}.
    \begin{itemize}
        \item \textbf{Combinatorial Auctions (CA), \cite{CA_01,CA_02,CA_03}:} Buyers and sellers are in multiple transactions on combinations of goods and services rather than individual ones. Although this would increase the buyer's utility and provide heterogeneous computing resources under complementary effects, there is a lack of discussion on maximizing social welfare. In addition, the seller and buyer rely on third-party auctioneers, which might raise concerns about sensitive data exposure. Furthermore, CA persists of computational complexity challenges, particularly winner determinations would be NP-hard.
        \item \textbf{Double Auctions (DA) \cite{DA_01,DA_02,DA_03}:} In DA, each seller presents prices to the auctioneer, and each buyer submits their bidding details. Then, the auctioneer matches buyers' bids with sellers' prices to find optimal deals. For instance, there are interactions in DA between multiple edge servers and users in the edge computing market via the auctioneer. However, the DA mechanism faces computational complexity challenges in larger-scale networks and privacy preservation during bid submissions, similar to CAs.
        \item \textbf{Combinatorial Double Auctions (CDA) \cite{CDA_01,CDA_02,CDA_03,CDA_04}:} This is a hybrid version of combining the advantages of CA and DA, allowing multiple sellers and buyers to trade on diverse item packages. Sellers submit price information for combinations of items, and buyers submit bidding information for desired combinations. The auctioneer then matches the requirements of sellers and buyers and determines the quantity and price of the package to be traded. In contrast, CDA privacy concerns arise from exposing sensitive bidding/price information for third-party auctioneers, resulting in a lack of data confidentiality. Moreover, challenges persist in handling the computational complexity of winner determination, which would be NP-hard optimization problems.
    \end{itemize}
As well, iterative combinatorial auctions have been employed to optimize resource allocation in mobile edge computing \cite{comb_auc_reversed, comb_auc_MEC, comb_auc_TCDA_MEC}. However, IoT devices are computationally limited and less motivated to engage in multidimensional decision-making by exploring time or frequency combinations instead of low-dimensional spatial frequency resource blocks. Additionally, the limited resource availability in dense networks encourages IoT entities to use simple bidding responses within the optimization task. Therefore, it is advantageous to experiment with iterative auctioning algorithms, such as simultaneous ascending auctions (SAA) \cite{Comb_SAA_auctions}, which involve bidders participating in individual bidding processes, particularly when resource block complementarities are minimal or absent. \emph{To the best of our knowledge, our proposed SAA-based auctioning approach is the first distributed resource allocation mechanism that combines lower computational complexity via a binary bidding process, specifically designed for the edge layer in dense IoT networks. In addition, the desired pipelining mechanism encourages IoT entities to calculate self-firing information on transmit power aligned with orthogonal spreading techniques, aiming at interference minimization while exploring the best resource block in the iterative period. The proposed framework distributes the computational overhead across individual IoT entities, thereby simplifying the otherwise massive centralized signal processing of multi-dimensional CSI and node hypotheses at the gateway in dense networks.}
\subsection{Contribution}
The primary contribution of our work is multifold. Specifically;
\begin{itemize}
    \item We consider massive and dense IoT networks ($K \geq N$) with limited communication resources $(N)$ relative to the number of IoT devices $(K)$ competing for them. We define a novel rationale for using auction theory-based scalable allocation mechanisms to develop policies enhancing user interactions between network gateways and nodes. We employ sealed-bid auctions for resource management to manage varying priorities and incomplete information available on inconsistent CSI among transceivers and unawareness of opponents' hypotheses.
    \item The independence of vendor-specific hardware and software constraints in IoT devices demands heterogeneous resource segments from the gateway. Therefore, we introduce space-time-frequency spreading (STFS) techniques for transmitting signals to reduce interference over the multi-access transmission channel while ensuring signal reliability and zero latency (\textit{explanation in Section \ref{STFS_discussion}}).
    \item We develop lightweight benchmark allocation mechanisms requiring minimal processing power to allocate optimal resource blocks for IoT nodes. This is achieved through a comprehensive mathematical exploration of traditional auction frameworks. Our proposed auction mechanism exhibits broader processing abilities, allowing the simultaneous distribution of computational complexity through sub-IoT clusters as illustrated in graphics (\figurename~\ref{Fig: SAA_node_interaction_both}). On the other hand, the state diagram motivates performance upgrades with additional graph-based signal processing blocks into the system to sharpen the self-predictability of IoT entities in future directions.
    \item Our proposed metric for optimality of the dispersion matrix activates STFS in IoT transmissions, ensuring proper regulation before compensating for external hardware and channel impairments. This approach reduces power consumption by preventing unnecessary collisions among transmit signals at the gateway while meeting the required channel throughput threshold, thereby minimizing network latency. In simulation results, our proposed mechanism outperforms the artificial neural network (ANN) model-based approach, establishing its superiority in terms of resource management with minimal energy budget for transmit signals, and robust in the presence of risky bidding nodes in the allocation mechanism. 
\end{itemize}

Our proposed SAA-based mechanism is advantageous for densely deployed IoT networks that must efficiently manage limited STFS resources. Despite high competition for STFS slot acquisition, the mechanism remains robust against nonstandard requests and ensures fair resource allocation. Additionally, IoT nodes can repurpose the power saved from inefficient transmissions for other signal processing and actuation tasks, thereby prolonging their primary battery life. On the gateway side, the system is capable of analyzing available STFS distributions and filtering out highly correlated ones before assigning them to IoT nodes. This minimizes signal collisions at the gateway, and facilitates the integration of new subscriptions into the network while maintaining high quality of service (QoS). Furthermore, spectrum resources are preserved, supporting the requirements for ultra-reliable and low-latency communication (URLLC).

Our proposed approach can be extended to wireless networks (5G, beyond-5G, and 6G), and will be highly opportunistic for mobile network operators, allowing them to provide feedback to mobile users and encourage active participation in transceiver operations. Meanwhile, the users can request specific STFS resource streams that best match their individual internal and external communication parameters, such as data transmission volume, channel strength, noise levels, acceptable latency, and more. Additionally, users will be able to make special requests for fair STFS resource allocation through bargaining strategies, especially in critical situations where they experience constraints in resources, like low battery levels, poor channel gain, or encounter the near-far problem, \cite{near_far_problem}. Since users respond to the prices set by the central base station, their confidential information remains undisclosed, thereby preserving user privacy. This platform promotes fair resource allocation by ensuring equal opportunities for all devices to participate in the decision-making process, mitigating monopolistic pricing control by the central base station.

\subsection{Organization}
In Section \ref{Sec: 2_system_model_problem}, we present the system model for the edge layer and formulate the resource allocation optimization problem using spreading techniques. Section \ref{Sec: 3_Auction_based_allocation} outlines the mathematical foundation of auction-based mechanisms for allocation policies, extending the sealed-bid approach to handle problem incompleteness. Section \ref{Sec: 4_Auction_types} explores conventional auction types as baselines for evaluation against the proposed framework outlined in Section~\ref{Sec: 5_proposed_approach_mSAA}, highlighting their unique properties and examining edge computing capabilities through dispersion metric optimizations.
Simulation results are provided in Section \ref{Sec: 6_results_discussion}, followed by a discussion aligned with resource allocation practices, concluding in Section \ref{Sec: 7_conclusion}. \footnote{\textit{Notation:} In this paper, bold upper case, lower case, and Calligraphy letters represent matrices, vectors, and sets or spaces respectively. $\mathbb{E}$, $\cross$, $\times$, $\dag$, $\Vert . \Vert$ and $|.|$ or $n(.)$ denote statistical expectation, the cross product, multiplication operator, Hermitian, the $L_{2}-$norm of a vector, and the cardinality of a set respectively. In addition, $|_{a}$ refer to \textit{with respect to $a$}, $\setminus k$ or $-k$ for \textit{without $k$}. $\rightarrow$ for \textit{maps to}, and $\perp$ refer to \textit{perpendicular} notations. Furthermore, $\mathbb{R}$, $\mathbb{C}$, $\cup$, $\textbf{I}_K$, $\textbf{0}_{K \cross N}$ denote the universal set of real numbers, universal set of complex numbers, the union of sets, $K \cross K$ identity matrix, $K \cross N$ null matrix respectively.}


\section{System Model and Problem Statement} \label{Sec: 2_system_model_problem}
\begin{figure*}[!t]
\centering
\includegraphics[width=0.99\linewidth]{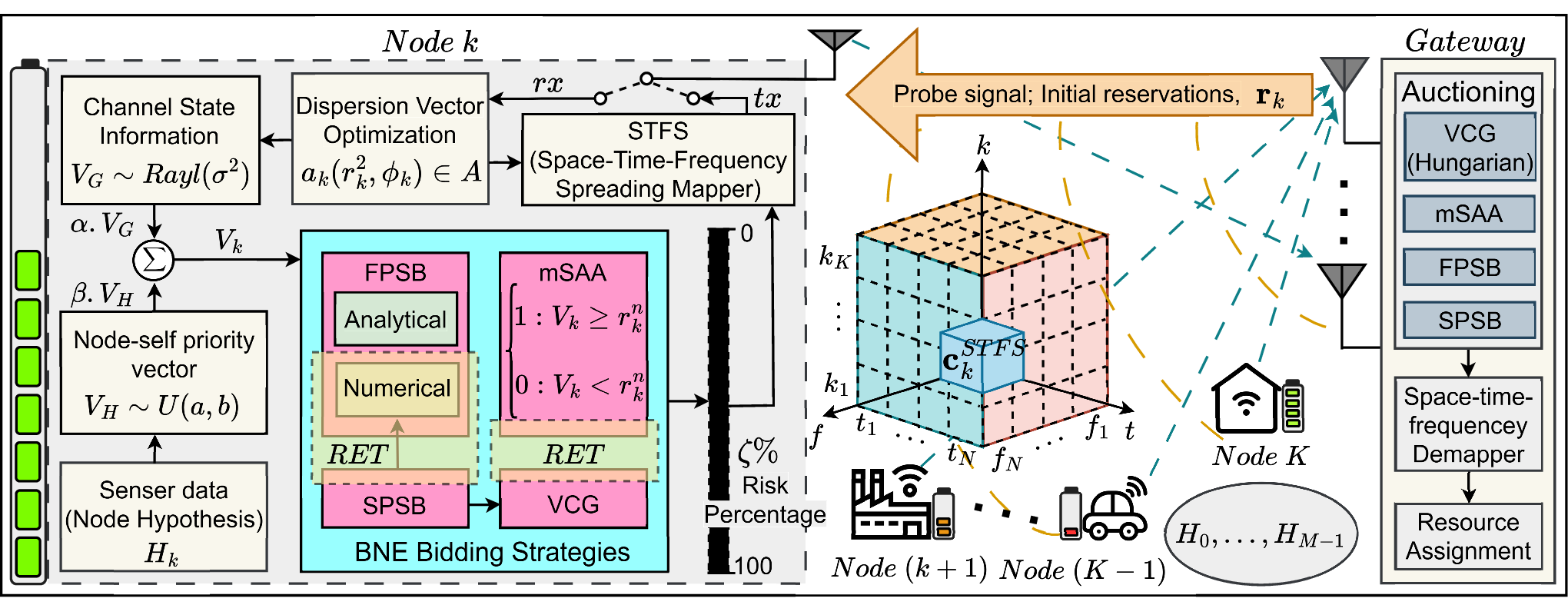}
\caption{Representation of Auction Game model-based simulators for the Co-Design systems of Resource allocation Optimization in Dense IoT network}
\label{Fig: System_model}
\end{figure*}
\subsection{Network Model}
Let us consider the massive $K$ number of IoT devices, each having unique attributes, are sending independent signals to their respective IoT gateways for uplink activities, as illustrated in \figurename~\ref{Fig: System_model}. These IoT devices are stationed in a fixed setup, capable of sensing various physical phenomena in their vicinity. Each IoT node $k \in \{1, \dots, K\}$ targets a specific resource slot $n \in \{1, \dots, N\}$. The received signal at the gateway at time $t$ can be formulated as,
\begin{align}
    y^{[n]}(t) = g_{k}(t) x_{k}^{[n]}(t) + \sum_{j=1, j\neq k}^{K} g_{j}(t) x_{j}^{[n]}(t) + w^{[n]}(t)
    \label{Eq: signal_model}
\end{align}
where $x_{k}(t) \in \mathbb{C}^{K \cross 1}$ and $g^{[n]}_{k} \in \mathbb{C}^{N \cross K}$ denote the intended transmitted signal from node $k$ and the corresponding channel vector for resource block $n$, respectively, both within the complex domain. These vectors follow a 2D complex bivariate Gaussian distribution to account for the effects of Rayleigh scattering resulting from small-scale path fading in the wireless medium\footnote{
The AWGN-based additive interference model in \eqref{Eq: signal_model} is adopted as a normalized abstraction to highlight allocation dynamics rather than detailed PHY-layer behavior. In practical IoT deployments, channel impairments such as frequency-selective fading, correlated multi-user interference, and non-Gaussian noise can be mapped into equivalent effective SNR shifts or variance scaling within the same framework. Moreover, the proposed dispersion optimization naturally extends to frequency-selective or correlated channels by compensating phase and gain distortions per sub-band, while the monotone price updates of mSAA remain agnostic to noise distribution. Thus, while our baseline model is simplified, the qualitative conclusions regarding convergence, surplus, and efficiency generalize to richer channel environments.}. The noise term $w^{[n]}(t)$ in \eqref{Eq: signal_model} is zero-mean complex Gaussian noise, where $\sigma_{w}^2$ is the average noise power.

In IoT networks with massive number of devices (where $ K \geq N $) and significant diversity, challenges arise not only in identifying any given resource block but also in precisely matching the most suitable slot at the gateway. The fundamental principle is that each resource segment $ n $ is designated for a single device $ k $ at most, using an allocation indicator $ c_{k}^{[n]} $ which is binary, valued either 0 or 1. This setup ensures that no device $ k $ is assigned multiple resources simultaneously. Therefore, the constraints are established as follows: the sum of allocations across all devices for each slot $ n $ does not exceed one ($ \sum_{k=1}^{K} c_{k}^{[n]} \leq 1 $), and similarly, each device $ k $ is restricted to one resource ($ \sum_{n=1}^{N} c_{k}^{[n]} \leq 1 $). These constraints apply for all $ n $ within the set of available slots $ \mathcal{N} $ and all $ k $ within the set of devices $\mathcal{K}$.

Let us define the data symbol vector of node $ k $ at time $ t $ as $ s_{k}(t) $. The complete set of these data symbols for all nodes is represented by the vector $ \textbf{s} = [s_{1}, \dots, s_{K}] $. This vector is generated based on the direct current (DC) transmission currents, $ \textbf{I}^{DC} = [I_{1}^{DC}, \dots, I_{K}^{DC}] $, and follows a circular symmetric complex Gaussian distribution. Specifically, each symbol $ s_{k} $ is distributed as $ s_{k} \sim \mathcal{C}\mathcal{N}_{\mathbb{C}}\left(0, P_{k}\propto{\big|I_{k}^{DC}}\big|^{2}\right) $, where $ P_{k} $ is the variance related to the square of the DC current of node $ k $. The data symbol $ s_{k} $ arose from binary phase shift keying (BPSK)-modulated signal transmitted by each node.
The covariance matrix for the vector $ \textbf{s} $ is the identity matrix $ \mathbf{I_{K}} $, ensuring that the transmission signals are independent and identically distributed (i.i.d). Consequently, the transmit signal from node $ k $ to the gateway, using resource block $ n $ at time $ t $, can be expressed as,
\begin{align}
    x_{k}^{[n]}(t) = c_{k}^{[n]} s_{k}(t); \quad \forall k \in \mathcal{K}, \forall n \in \mathcal{N}.
    \label{Eq: tx_signal}
\end{align}

\subsection{Space-Time-Frequency-Spreading (STFS)} \label{STFS_discussion}
The architecture depicted in \figurename~\ref{Fig: System_model} outlines a data mapping technique for the data stream $s_{k}$ of node $k$, employing a pre-calculated metric known as the dispersion vector $a_{k} \in \textbf{A}$. This vector is applied before the data is transmitted over combinations of space, time, and frequency. The detailed configuration of the dispersion matrix is represented as $\textbf{A}~\in~\mathbb{C}^{K \cross N_{T} \cross N_{F}}$, where  $N_{T}=n(\mathcal{N}_{T})$ and $N_{F}=n(\mathcal{N}_{F})$ are the numbers of time and frequency slots available within the total set of resource slots $\{\mathcal{N}_{T} \cup \mathcal{N}_{F}\}\subseteq\mathcal{N}$, and each IoT node in set $\mathcal{K}$ is assigned specific slots from these subsets. The optimal dispersion vector is refined by incorporating prior compensation for channel characteristics in the complex plane, enhancing power efficiency and synchronization between time and frequency before transmitting signals from each IoT node, (\textit{explanation in Section \ref{sec_sub: dispersion_mat_opt_A}}) \cite{STFS_indi}. 

In real-world IoT networks, the desired STFS resource block depends on the required bandwidth of the data stream and the individual hypothesis of the actuated data vector of IoT nodes. Although IoT nodes would be able to select a suitable STFS slot compatible with existing intrinsic requirements, the transmit signal is expected to deviate from the pre-reserved STFS slot for ambient effects. The assignment of a multilevel channel throughput threshold that compromises various receive signal strength indications (RSSI) for each node could require context-aware STFS allocation. Moreover, selecting the optimal STFS slot is challenging for low-SWaP IoT nodes with lower processing overhead, as they must avoid interference-prone areas within the corresponding STFS resource cube.

STFS constructs a three-dimensional mapping framework using blocks from the matrix $\textbf{A} \in \mathbb{C}^{K \times N_{T} \times N_{F}}$. The flexibility of the STFS allocation parameter $c_k$ allows for dynamic resource assignment sensitive to time and frequency demands. For the $k$th device, the STFS modifies the transmit signal as follows:
\begin{align}
    x_{k}^{[n]}(t) \Big|_{\textbf{a}}  = a_{k}(t). x_{k}^{[n]}(t); \quad \forall k \in \mathcal{K}, \forall n \in \mathcal{N}
    \label{Eq: Tx_signal_STFS}
\end{align}
where $p_k = \textbf{a}_{k}^{\dag} \cdot \textbf{a}_{k}$, with $\textbf{a}_{k} \in \mathbb{C}^{l \times 1}$ and $l$ represents the length of the symbol stream. The transmit power for node $k$ must be sufficient to meet the gateway's RSSI requirements. The average power constraint $P$ is then defined to ensure the power used does not exceed available resources:
\begin{align}
    \sum_{n=1}^{N} \mathbb{E} \left(\Big \Vert x_{k}^{[n]}(t) \Big \Vert_{\textbf{a}}^{2} \right) \leq P; \quad \{c_{k}^{[n]} \perp p_{k}^{[n]}\} \nonumber \\ c_{k}^{[n]} \in [0,1], \forall k \in \mathcal{K}, \forall n \in \mathcal{N} 
    \label{Eq: tx_signal_power_avg}
\end{align}
While Eq.~\eqref{Eq: tx_signal_power_avg} specifies the average transmit-signal power constraint, it models only the radiated RF component. In practice, IoT devices also consume significant baseband and RF front-end power, typically around $40\%$ of the total node energy budget. To reflect this, the effective device-side power consumption can be expressed as $P_{\mathrm{tot}} = P_{\mathrm{tx}} + P_{\mathrm{bb}}$ where $P_{\mathrm{tx}}$ corresponds to the constrained transmit power in \eqref{Eq: tx_signal_power_avg} and $P_{\mathrm{bb}} \approx 0.4 P_{\mathrm{tot}}$ accounts for circuitry overhead. This adjustment ensures that the subsequent energy-efficiency analysis better aligns with practical IoT device hardware. The above equation sets an upper limit on the power variance $P_k \leq P$ using the gain from the optimal STFS block $a_k$. Importantly, while the allocation indicator $c_{k}^{[n]}$ may influence the selection of $a_{k}^{[n]}$, the power used $p_{k}^{[n]}$ by each node $k$ for resource $n$ remains independent of $c_{k}^{[n]}$.

\subsection{Problem Formulation}

When diverse resource allocation protocols are introduced, the interoperability between a large number of IoT devices can deviate from expected execution patterns, moving away from deterministic behaviors. The issue of resource scarcity, particularly when the number of devices exceeds available resources, i.e. $K > N$, enhances competitive interactions among devices. This competition often leads to signal collisions at the gateway due to the delay spread of the transmitted signals, which can exceed the coherence time of the channel at higher frequency stages, causing multipath fading effects. A notable issue arises when a signal designated for a specific resource block $n$ from an IoT device $k $, interferes with signals from other devices that are also assigned to the same resource block $r_{n}^{STFS}$ at the vicinity of $n$, leading to a potential reduction in channel throughput. This throughput for each device and resource block, assuming no other interferences, can be quantified by:
\begin{align}
    \gamma_{k}^{[n]}& = B\log_2 \left(1+ \frac{\Big\Vert a_{k}g_{k}x_{k}^{[n]}\Big\Vert^2}{\Big\Vert \sum_{j=1, j\ne k}^{K}a_{j} g_{j} x_{j}^{[n]}\Big\Vert^2+{\sigma_{w}}^2}\right)
    \label{Eq: throughput_x}
\end{align}
where $B$ is the normalized bandwidth based on the specific IoT attributes and surrounding conditions, $a_k$ and $g_k$ are the dispersion and channel gain vectors for device $k$, and $\sigma_{w}^2$ is the noise power. Furthermore, the term for interference power benefits the phase details among summation of interference signals making a proper resource arrangement utilizing $c_{k}^{[n]}$ and executing $a_{k}$ may cause to enhance the channel throughput of the received signal at the gateway, fired from IoT device $k$. Ultimately, the formal building blocks for the adaptive resource allocation optimization problem are as follows.
\begin{subequations}\label{Eq: Opt_problem_main}
\begin{IEEEeqnarray}{cl}
\underset{\mathbf{C},\mathbf{A},\mathbf{P}}{\mbox{minimize}}&~~ \Bigl\{ \frac{1}{K} \sum\limits_{k\in\mathcal{K}} p_{k}(\mathbf{C},\mathbf{A}) \Bigl\} \in \mathbf{P} \label{Eq: objective_fn}
\\ \mbox{subject to}&~~ \gamma_{k}(\mathbf{C},\mathbf{A},\mathbf{P}) \geq \Gamma^{th}_{k}; \quad c_{k}^{[n]}\in \mathbf{C}, a_{k}^{[n]}\in \mathbf{A} \label{Eq: gamma_th}
\\&~~\quad p_{k}^{\min} \leq p_{k}\leq p_{k}^{\max} \in P \label{Eq: power_bounds}
\\&~~\quad\eqref{Eq: tx_signal}, \eqref{Eq: Tx_signal_STFS}, \eqref{Eq: tx_signal_power_avg}, \eqref{Eq: throughput_x}; \quad \forall k\in \mathcal{K}, \forall n\in \mathcal{N}. \label{Eq: rest_equations}
\end{IEEEeqnarray}
\end{subequations}
The objective function \eqref{Eq: objective_fn} represents the average transmit power of uplinks in the IoT pool with the regulation flexibility upon STFS indicator $\mathbf{C}$ and dispersion metric $\mathbf{A}$. The power minimization problem \eqref{Eq: power_bounds} is constrained with a given acceptable lower bound of channel throughput value, $\Gamma^{th}_{k}$ of node $k$ as in \eqref{Eq: gamma_th} for better QoS. The rest of the constraints, \eqref{Eq: rest_equations} act as supportive agents for the execution mechanism and the completion of the optimization problem. 

\section{Auction-based Resource Allocation} \label{Sec: 3_Auction_based_allocation}
Each IoT node will have unique characteristics in the resource allocation framework, considering the internal self-requirement, such as the amount of battery capacity, the size of the uplink data packet, and the urgency of reaching the endpoint. Additionally, there are external impacts to achieve with available STFS resource blocks and correlation with the transmit signal, potential for guarding against interference, etc. Although IoT entities have knowledge about self-impact parameters, it is challenging to acquire external details owing to the larger-scale dynamics in the network. Therefore, competitive IoT devices demand premium resource blocks in the dense IoT network following the built-in game model for resource allocation. Furthermore, the prior joint distribution of incomplete parameters supports taking Bayesian strategies rewarding the equilibrium allocation state. Nevertheless, the performance of the Bayesian-only strategies approach may be inefficient for low-SWaP IoT entities and requires a distributed computational architecture, although it adheres to the fundamental characteristics of Bayesian game models. Hence, auction theory, a branch of Bayesian game theory capable of capturing user interactions during a single execution of a federated algorithm, serves as an ideal approach for optimization use cases.

In the context of auction-based resource allocation for IoT networks, we address the optimization challenge where the nature of the problem changes due to the unique characteristics of each IoT device and their interaction with the network's centralized gateway. This interaction often introduces complexities not suited to straightforward optimization due to the dynamic nature and incomplete information about each node's state, making the problem non-convex and more complex to solve through traditional methods. Auction theory offers a robust framework for allocating resources in dense IoT environments where competition for resources ($K \geq N$) can lead to inefficiencies and conflicts. It provides a structured approach to distribute resources fairly and efficiently among multiple competing nodes (buyers) and the gateway (seller), leveraging the principles of game theory to navigate through sub-optimization challenges.

The optimization problem, formalized in the main equation \eqref{Eq: Opt_problem_main}, aims to minimize power usage while ensuring quality service by maintaining a minimum throughput ($\Gamma^{th}_{k}$) for each device. In an auction setting, IoT devices (buyers) aim to minimize their energy consumption (cost) while competing for network resources, which are efficiently allocated by the gateway (seller) through auction mechanisms. The goal is to find a power strategy and resource block allocation that aligns with each device's operational characteristics and network conditions. To optimize resource usage further, we consider a sub-optimization problem that focuses on maximizing the surplus for each device:
\begin{subequations}\label{Eq: sub_opt_surplus}
\begin{IEEEeqnarray}{cl}
\underset{\mathbf{C},\mathbf{A},\mathbf{P}}{\mbox{maximize}}&~~ \sum_{k=1}^{K} S_{k}(\mathbf{C},\mathbf{A},\mathbf{P})
\\ \mbox{subject to} &~~ S_{k}(\mathbf{C},\mathbf{A},\mathbf{P}) = P^{tot}_{k} - p_{k} \label{Eq: surplus_equation}
\end{IEEEeqnarray}
\end{subequations}
where $S_{k}(\mathbf{C},\mathbf{A},\mathbf{P}) = P^{tot}_{k} - p_{k}$ reflects the surplus each device aims to maximize by reducing power consumption, under the constraint that it doesn’t compromise the required service quality. This approach promotes energy efficiency and prolongs battery life, crucial for mobile and remote IoT applications. Conversely, the gateway's objective is to maximize the aggregate channel throughput, which can be seen as maximizing its revenue from selling resource slots to the highest bidders among the IoT devices:
\begin{subequations}\label{Eq: sub_opt_revenue}
\begin{IEEEeqnarray}{cl}
\underset{\mathbf{C},\mathbf{A},\mathbf{P}}{\mbox{maximize}}&~~ \sum_{n=1}^{N} R_{n}(\mathbf{C},\mathbf{A},\mathbf{P})
\\ \mbox{subject to} &~~ R_{n}(\mathbf{C},\mathbf{A},\mathbf{P}) = \gamma_{n\leftarrow k} \label{Eq: revenue_equation}
\end{IEEEeqnarray}
\end{subequations}
where, $R_{n}(\mathbf{C},\mathbf{A},\mathbf{P}) = \gamma_{n\leftarrow k}$ represents the throughput for each resource block $n$, maximized through optimal allocation. This optimization ensures that the network operates at peak efficiency, balancing the demand and supply of network resources effectively.

\subsection{Sealed-Bid Auction for Dense IoT Networks}
Sealed-bid auctions provide a robust mechanism for distributing limited resources among competing IoT devices (buyers) and a centralized gateway (seller), each with its own set of private values and information. In these auctions, each device submits a bid for the resources without knowledge of the bids from other devices, making this method highly suitable for environments where sharing detailed bid information might lead to strategic bidding or collusion. In this context, the exact valuation for the bid, each device places on a given resource can be private, influenced by individual operational requirements, future resource needs, or even strategic considerations. These valuations are not always fully known to other devices or the gateway, introducing uncertainty and complexity in the auction process. The mathematical model for a sealed-bid auction in the above setting can be expressed as,
\begin{align}\label{Eq: auc_game_model}
    \mathcal{G_A} \triangleq \Big \langle \hspace{-0.6mm} \mathcal{K}, \mathcal{N}, (\mathcal{V}_{k}, \mathcal{F}_{k})_{k \in \mathcal{K}}, \mathcal{B}, \mathcal{Q}, \mathcal{U} \hspace{-1mm} \in \hspace{-1mm} \{(\mathcal{S}_{k})_{k \in \mathcal{K}} \hspace{-0.5mm} \cup \hspace{-0.5mm} (\mathcal{R}_{n})_{n \in \mathcal{N}}\} \hspace{-0.6mm} \Big \rangle
\end{align}
where, $\mathcal{K} =\{1,\ldots,k, \ldots, K\}$ is the set of IoT devices (buyers), $\mathcal{N} =\{1,\ldots,n, \ldots, N\}$ is the set of STFS resource slots or selling objects at the IoT gateway, $\mathcal{V}_{k} \subseteq \mathbb{R}: v_{k}$ is the anticipated the private valuations of node $k\in \mathcal{K}$, $\mathcal{V}^{\mathcal{K}}\coloneqq\mathcal{V}_{1} \ldots \times \mathcal{V}_{K}$ is the set of private valuations. The cumulative distribution function for the set of private valuations $\mathcal{V}_{k}$ is $\mathcal{F}_{\mathcal{V}}(v)=\int_{\underline{v}}^{v}f_{\mathcal{V}}(t)dt; \quad \forall \{ \underline{v}, v \} \in \mathbb{R}^{+}_{0} $. The joint probability distribution of multi-dimensional mutual prior beliefs is modeled as, $f_{GH}(g,h)$ employing marginal distribution $f_{G}(g) = \int_{-\infty}^{\infty} f_{GH}(g,h) dh$, and  $f_{H}(h) = \int_{-\infty}^{\infty} f_{GH}(g,h) dg$ for CSI and individual hypothesis vector respectively. $\mathcal{B}=\{b_{1},\ldots,b_{k}, \ldots, b_{K}\}$ is the action set of each IoT node and $p: [0, \infty)^{\mathcal{K}}\rightarrow\Delta(\mathcal{K})$ denotes the function aligning each bidding vector $\textbf{b}\in[0, \infty)^{\mathcal{K}}$ with a distribution that associates winning IoT devices to recognized resource objects and $~\Delta(\mathcal{K})\coloneqq\{ x \in [0,1]^{\mathcal{K}}: \sum_{k \in \mathcal{K}}x_{k}=1 \} $ is the combination of possible probability distributions over the set of nodes $\mathcal{K}$. And $\mathcal{Q}: \mathcal{K} \cross [0, \infty)^{\mathcal{K}}~\rightarrow~\mathbb{R}^{\mathcal{K}}$ is the cost function associating with payment details for each winning node $k^{*} \in \mathcal{K}$ and the gateway earns $q_{n}$ for served resource slot $n \in \mathcal{N}$. The payoff or utility function of the auction models is indicated by $\mathcal{U}: \mathcal{B}~\rightarrow~\mathbb{R}$ associating the bidding distribution $p$ where $u_{k^{*}}(\textbf{b})$ is the utility of winner node $k^{*}$ and winning probability $p_{k^{*}}(\textbf{b}=\{ b_{1}, \dots, b_{K} \})$. Every leading IoT node $k^{*}$ consumes the sum cost: $Q_{k}(k^{*}; b_{1}, \dots, b_{K})$ and proportionally maps with revenue of $n^{th}$ resources at the gateway, $(k^{*}\rightarrow n)$. Hence, the generalized utility space, $\mathcal{U_{\mathcal{S},\mathcal{R}}}$ of surplus:~$\mathcal{S}_{\mathcal{K}}$ and revenue:~$\mathcal{R}_{\mathcal{N}}$ for the IoT pool and gateway can be expressed as;
\begin{align}\label{Eq: utility_S_R}
    \begin{dcases}
        \mathcal{S}_{\mathcal{K}} = \sum_{k \in \mathcal{K}} \left(p_{k}(\textbf{b})v_{k} - \sum_{k^{*} \rightarrow n \in \mathcal{N}} p_{k^{*} }(\textbf{b})Q_{k}(k^{*}\rightarrow n;\textbf{b})\right)
        \\
        \mathcal{R}_{\mathcal{N}} = \sum_{k^{*} \rightarrow n \in \mathcal{N}} p_{k^{*} }(\textbf{b})Q_{k}(k^{*}\rightarrow n;\textbf{b}) 
    \end{dcases}
\end{align}
Thus, the mathematical discussion for sealed bid auction fits perfectly with Harsanyi's game model \cite{Harsanyis_GT} for the problems with incomplete information. Consequently, the equilibrium state of the game model can be obtained utilizing the conditional probability approach called Bayesian strategies. In a sealed-bid auction-based resource allocation,  the pure bidding strategy $\beta_{k}$ of node $k$ is a measurable function. Each number $y \in \left[0, \infty \right)$, the set of $g^{-1}(\left[0, y \right])= \{x \in X: g(x) \leq y \}$ is a measurable set $\Rightarrow$ A real-valued function $g: X \rightarrow \left[0, \infty \right)$ is \textit{measurable}; $\forall X \subseteq \mathbb{R}$.

Let us define $\beta = (\beta_1, \dots, \beta_K)$ where each $\beta_k: [0, \infty) \rightarrow [0, \infty)$ represents the bidding strategy of IoT device $k$. The vector $\beta_{-k}(b_{-k})$ comprises the strategies of all nodes except $k$, reflecting their respective bids and valuation types. The expected utility or surplus $u_k(\beta; v_k)$ for device $k$, which depends on both the strategy vector $\beta$ and $k$'s private valuation $v_k$, is given by \eqref{Eq: expected_utility}.
\begin{figure*}
\begin{align} \label{Eq: expected_utility}
        u_{k}\left(\beta; v_{k}\right) \coloneqq \bigintssss_{\mathcal{V}_{-k}} \biggl( p_{k} \Bigl(\beta_{k}(v_{k}), \beta_{-k}(b_{-k}) \Bigr)v_{k} - \sum_{k_{*} \in \mathcal{K}} p_{k_{*}} \Bigl(\beta_{k}(v_{k}), \beta_{-k}(b_{-k}) \Bigr). Q_{k}\Bigl(k_{*}; \beta_{k}(v_{k}), \beta_{-k}(b_{-k}) \Bigr) \biggr) dF_{-k}(b_{-k}).
\end{align}
\noindent\makebox[\linewidth]{\rule{0.84\paperwidth}{0.4pt}}
\end{figure*}
There, $\mathcal{V}_{-k} = \times_{j \neq k} \mathcal{V}_j$ represents the combined private valuation spaces of all nodes except $k$, and $F_{-k} = \times_{j \neq k} F_j$ is the joint cumulative distribution function of their multidimensional prior beliefs. This formulation captures how $u_k(\beta; v_k)$ depends solely on $k$'s strategy and the responses of other nodes, aligning with their private values and strategic decisions. The BNE is achieved when no participant $k$ can unilaterally change their bid to improve their outcome, formalized as:
\begin{align}\label{Eq: BNE}
    u_{k}(\beta^{*};v_{k}) \geq u_{k}(b_{k},\beta^{*}_{-k};v_{k});\forall b_{k} \in [0,\infty), \forall v_{k} \in \mathcal{V}_{k}, \forall k \in \mathcal{K}.
\end{align}
At this equilibrium, each IoT device's bidding strategy $\beta_k^*(v_k)$ maximizes its expected utility given the strategies of others, indicating optimal resource allocation based on private valuations within the network. This setup highlights the complex interdependencies among devices' private valuations and the need for a cooperative protocol to address collective optimization challenges beyond individualistic approaches.

\section{Auction types} \label{Sec: 4_Auction_types}
In this section, we discuss the sealed-bid auction model and its application in IoT networks, focusing on the valuation space $\mathcal{V}_k$ of each IoT device $k$ within the network. Each device's valuation space comprises two primary components: the individual hypothesis $v_{h}^k$, and the channel state information (CSI), $v_{g}^k$. These elements are crucial for devices when competing for the optimal STFS slot. The individual hypothesis follows a uniform distribution, $\mathcal{V}^H \sim \mathbb{U}(a, b)$, with a probability density function (PDF) given by $f_H(h) = \frac{1}{b-a}$, for $v_h \in [a, b]$. This uniform distribution represents an equal likelihood of any priority level between $a$ and $b$.

Moreover, the CSI part of the valuation, $v_{g}^k$, considering the link budget and mobility influences, follows a Rayleigh distribution, $\mathcal{V}^G \sim \text{Rayl}(\sigma^2)$, with a PDF $f_G(g) = \frac{v_g}{\sigma^2} e^{-\left(\frac{v_g^2}{2\sigma^2}\right)}$ for $v_g \geq 0$. This distribution is used to model the small-scale path-fading effects on the channel. Combining these distributions, the overall valuation space for each node is expressed as $\mathcal{V} = \alpha \mathcal{V}^H + \beta \mathcal{V}^G$, where $\alpha$ and $\beta$ are numerical weights that adjust the influence of hypothesis and channel strength in the valuation model, depending on the specific IoT application requirements. The joint probability density function for the valuation space, integrating these two components, is given by $f_{\mathcal{V}^{G}, \mathcal{V}^{H}}(v_g, v_h) = \frac{1}{(b-a)} \cdot \frac{v_g}{\sigma^2} e^{-\left(\frac{v_g^2}{2\sigma^2}\right)}$. The cumulative distribution function (CDF) of the entire valuation space, integrating over all possible values of $v_g$ and $v_h$, can be approximated by:
\begin{align}
    \begin{split} \label{Eq: CDF_valuation_distribution}
        F_{\mathcal{V}}(v) &= \int_{-\infty}^{\infty}\int_{-\infty}^{\infty}f_{\mathcal{V}^{G}, \mathcal{V}^{H}}(v_{g},v_{h})dv_{g} dv_{h}\\
        &;v_{h} \in [a,b], \quad v_{g} \geq 0 \textit{ and}, \quad \mathcal{V}=\alpha \mathcal{V}^{H} + \beta \mathcal{V}^{G}\\
        F_{\mathcal{V}}(v)  &= \hspace{-1mm} 1 + \hspace{-1mm} \frac{\beta \sigma}{(b-a)\alpha} \hspace{-0.1mm} \sqrt{\frac{\pi}{2}} \hspace{-1mm} \left[ erf \hspace{-1mm} \left(\frac{v-\alpha b}{\sqrt{2}\beta\sigma}\right)-erf \hspace{-1mm} \left(\frac{v-\alpha a}{\sqrt{2}\beta\sigma} \right) \hspace{-1mm} \right]\\
        &;erf(b)-erf(a) = \frac{2}{\sqrt{\pi}}\hspace{-1mm}\int_{a}^{b}e^{-u^{2}} du,\quad u=\hspace{-1mm}\frac{\left(z-\alpha.v_{h}\right)}{\sqrt{2}\beta\sigma}
    \end{split}
\end{align}
where $\text{erf}$ is the error function, reflecting the integration of the Gaussian components of the valuation distributions. 
According to the objective function in \eqref{Eq: utility_S_R}, the optimization variable is identified as the bidding strategy vector $\mathcal{B}$ which is related to the payment cost $\mathcal{Q}$ in \eqref{Eq: auc_game_model}, for optimal resource demanding from each IoT node, $k \in \mathcal{K}$ to the preferred STFS slot, $n \in \mathcal{N}$. Let's consider different sealed-bid auction models to evaluate the bidding function and performance themselves.

\subsection{Traditional auctions}
In this section, we explore classical single-object ascending and sealed-bid auctions to propose optimal resource assignments in dense IoT networks. To maintain fairness in resource allocation, it is stipulated that each node can receive at most one STFS slot in a single auction. Viewing the multi-objective auction as a series of independent single-object auctions provides a clearer understanding of the underlying auction fundamentals, which are crucial benchmarks for assessing the performance of more complex auction mechanisms. In ascending bid auctions, the highest bidder typically secures the desired resource. If multiple nodes place the same highest bid, a fair lottery determines the assignment of the STFS slot. This scenario, although unlikely in continuous probability distributions, ensures that each auction outcome is determined justly\footnote{The likelihood of observing multiple players with identical highest bidding strategy vectors and the same probabilities for their self-valuations in a continuous probability distribution is negligible.}.

\subsubsection{Second-Price Sealed-Bid Auction (SPSB)} 
Also known as a Vickrey auction, SPSB is a fundamental model for understanding optimal bidding strategies or BNE for each node, as established in the following lemma~\ref{Thm: BNE_SPSB}, (proof: P91-94, theorem~$4.15$ in \cite{maschler_solan_zamir_2013}).
\begin{lemma} \label{Thm: BNE_SPSB}
In a second-price sealed-bid auction, the strategy $b_{k}=v_{k}$ weakly dominates all other strategies, $b~\in~\mathcal{B}, v~\in~\mathcal{V}, k~\in~\mathcal{K}.$ 
\end{lemma}
The winner node should pay the bidding amount of the second-highest bid as the price to the gateway. In other words, the IoT device, $k \in \mathcal{K}$ with the highest valuation $v_{k}$ is assigned to a given STFS resource segment, $n\in \mathcal{N}$ and $c_{[k]}^{n}=1$ taking into account the individual hypothesis $v_{h}^{k}$ and CSI $v_{g}^{k}$. Then, the gateway accepts the channel throughput $\gamma^{n}_{k}$ of $k^{th}$ node for fair resource $n$, which exactly matches with the channel throughput threshold $\Gamma^{th,n}_{k^{'}}$, of the IoT device $k^{'}$  that has the second highest bid $b_{k^{'}}$ (mirror the characteristics of node $k^{'}$-self) for the same STFS slot $n$, $(q_{k}^{[n]}=b_{{k^{'}}}^{[n]})$. Therefore, IoT device $k$ has the opportunity to decrease the power consumption of the transmit signal to satisfy only with second best SINR instead of the self-highest SINR for STFS slot $n$, $(\Gamma^{th,n}_{k^{'}} \leq \gamma^{n}_{k})$. Additionally, resource sharing in accordance with the flow of priority could further enhance guaranteed fair resource allocation. 

\subsubsection{First Price Sealed-Bid Auction (FPSB)}
In an ascending FPSB auction, IoT nodes submit sealed bids $\mathcal{B}$, and the node that submits the highest bid is awarded the STFS resource slot $n$ and pays the exact bid amount to the gateway. Here the payment $q_{k}^{[n]}=b_{k}$ from winner node $k$ to the gateway for $n^{th}$ resource, represents the amount of harnessing energy required for transmission to meet the given channel throughput threshold $\Gamma^{n,th}_{k}$. It also reflects the fairness factor concerning the suitability of the assigned resource slot relative to the self-hypothesis, similar to the SPSB auction. However, based on these rules, it is evident that nodes will be disinclined to bid true valuations themselves due to receiving zero profit or surplus as explained in \eqref{Eq: surplus_equation}. Lower or zero surplus means the less or null power saving $\left(S_{k}\left(v_{k}^{g}\right) \rightarrow 0\right)$, and this, combined with reduced consideration for the hypothesis of neighboring devices $\left(S_{k}\left(v_{k}^{h}\right) \rightarrow 0\right)$ may negatively affect for the overall efficiency of the network. Therefore, the BNE of the bidding strategy in the FPSB auction should be lower than the actual valuation $v_{k}$ of each node $k$ to make a positive profit potentially. To find the BNE in the FPSB auction, let's consider that each node utilizes a bidding function $\mathbf{b}$, which is strictly increasing, continuous, and differentiable for the signals in valuation. And, if node $k \in \mathcal{K}$ uses identical bidding strategies $b_{k}=\mathbf{b}(v_{k})$ from valuations $v_{k}$ along a symmetric distribution, then the expected utility of node $k$ for $n^{th}$ STFS slot, as a function of bid $b_{k}$ and \eqref{Eq: expected_utility} is, \footnote{$b_{-k}=\{b_{1},\ldots, b_{k-1}, b_{k+1},\ldots, b_{K}\}$ is the bidding vector of IoT pool except $k^{th}$ node, and the losing nodes in the auction receive nothing results in a null vector for utilities themselves, $S_{(k^{lose})}=\textbf{0}$.}
\begin{align} \label{Eq: expected_utility_fpsb}
    \begin{split}
        \mathbb{E}\left[U_{k}(b_{k}, b_{-k}, v_{k})\right] &= S_{(k^{win})}.\Pr (\text{$k^{win}$})  + S_{(k^{lose})}.\Pr (\text{$k^{lose}$})\\
        &= (v_{k}-b_{k}). \hspace{-0.4mm} \Pr \left[(b_{j} \hspace{-0.5mm} = \hspace{-0.5mm} \mathbf{b}(v_{j}) \leq b_{k}, \forall j \neq k)\right]\\
        &= (v_{k}-b_{k}). {F_{\mathcal{V}}}^{K-1}\left[(\mathbf{b}^{-1}(b_{k})\right]
    \end{split}
\end{align}
where, $\Pr [v_{j} \leq \mathbf{b}^{-1}(b_{k})]={F_{\mathcal{V}}}^{K-1}[\mathbf{b}^{-1}(b_{k})]$\footnote{Here, the node valuation vectors are independent and the prior cumulative distributive function $F_{\mathcal{V}}$ of bidding strategy follows the rule of, $\Pr\left(\cap_{\forall j \in \{\mathcal{K} \setminus k\}} b_{j} \leq b_{k}\right) = \prod_{j \in \mathcal{K} \setminus k} \Pr \left(b_{j} \leq b_{k} \right)$ for the winning probability of node $k$ with winning bid $b_{k}$.}$;~\forall~j~\in~\mathcal{K}\setminus k$. Then, the $\max_{b_{k}} (v_{k}-b_{k}). {F_{\mathcal{V}}}^{K-1}\left[(\mathbf{b}^{-1}(b_{k})\right]$ expected utility maximization problem from \eqref{Eq: BNE}, derives the formula for BNE as in \eqref{Eq: BNE_FPSB_general} for IoT nodes concerning $b_{k}$, (proof: P3-5, section $1.3$ in \cite{Auc_theory_stanford}). 
\begin{figure*}
\vspace{-7mm}
\begin{align}
    b(v^{1}, \dots, v^{L})=\mathbf{f}_{v}(v^{1}, \dots, v^{L})-\frac{\bigintss_{\textbf{$\underline{v}^{1}$}}^{\textbf{$v$}^{1}} \dots \bigintss_{\textbf{$\underline{v}^{L}$}}^{\textbf{$v$}^{L}}  {\mathbf{f}_{v}}^{K-1}\left(\prod_{l=1}^{L} F_{\mathcal{V}_{l}}(t_{l}) \right) \,dt_{1} \dots dt_{L}}
    {{\mathbf{f}_{v}}^{K-1}\left(\prod_{l=1}^{L} F_{\mathcal{V}_{l}}(v_{l}) \right)} \label{Eq: BNE_FPSB_general}\\
    b\left(v_{h,g}=\alpha v_{h}+\beta v_{g}\right) = \left(\alpha v_{h}+\beta v_{g}\right) - \frac{\bigintss_{\underline{v}_{h,g}}^{v_{h,g}} \left(1 + \frac{\beta \sigma}{(b-a)\alpha}\sqrt{\frac{\pi}{2}}\left[ erf\left(\frac{t-\alpha b}{\sqrt{2}\beta\sigma}\right)-erf\left(\frac{t-\alpha a}{\sqrt{2}\beta\sigma} \right)\right]\right)^{K-1}dt} {\left(1 + \frac{\beta \sigma}{(b-a)\alpha}\sqrt{\frac{\pi}{2}}\left[ erf\left(\frac{\alpha v_{h}+\beta v_{g}-\alpha b}{\sqrt{2}\beta\sigma}\right)-erf\left(\frac{\alpha v_{h}+\beta v_{g}-\alpha a}{\sqrt{2}\beta\sigma} \right)\right]\right)^{K-1}} \label{Eq: BNE_FPSB_Vh_Vg}
\end{align}
\vspace{-1mm}
\noindent\makebox[\linewidth]{\rule{0.84\paperwidth}{0.4pt}}
\vspace{-3mm}
\end{figure*}

\begin{figure*}[t!]
\vspace{-4mm}
\centering
$\begin{array}{ccc}
\includegraphics[width=0.3\linewidth, trim={0mm 0mm 10mm 7mm},clip]{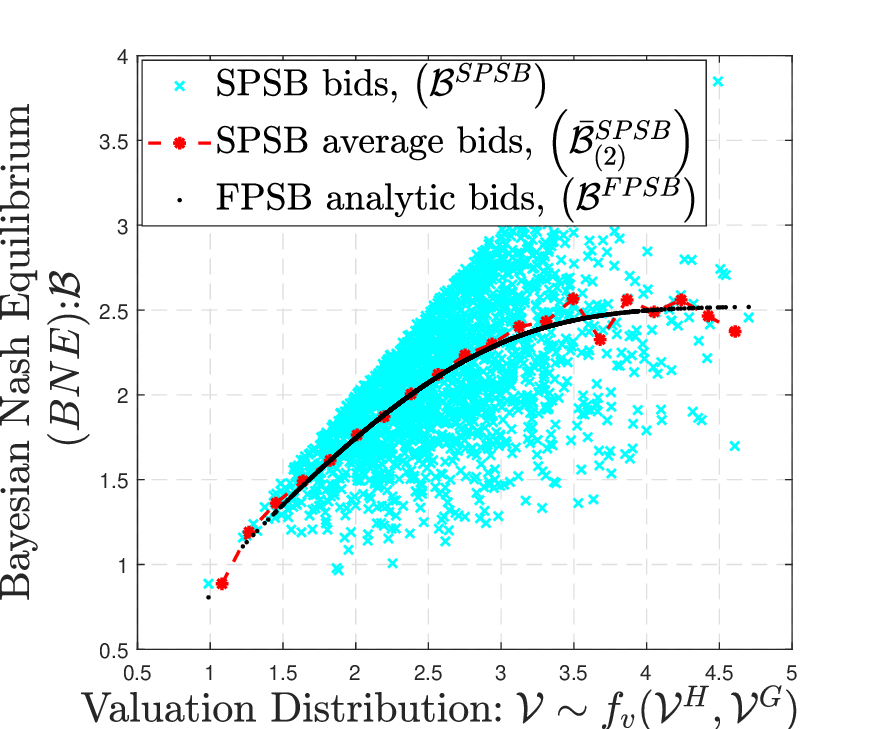} &
\includegraphics[width=0.33\linewidth, trim={0mm 1mm 10mm 7.5mm},clip]{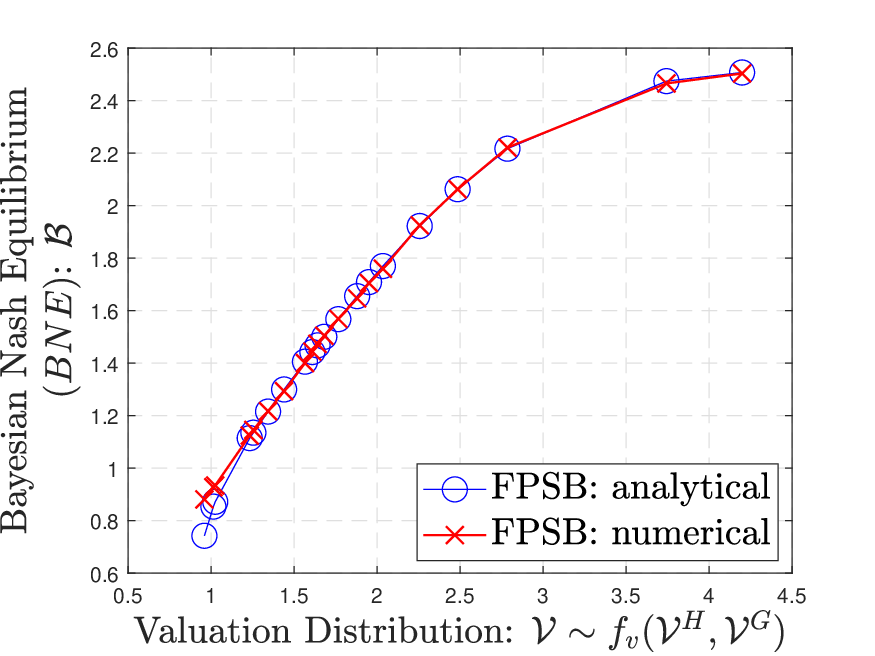} &
\includegraphics[width=0.32\linewidth, trim={1mm 1mm 9mm 5.3mm},clip]{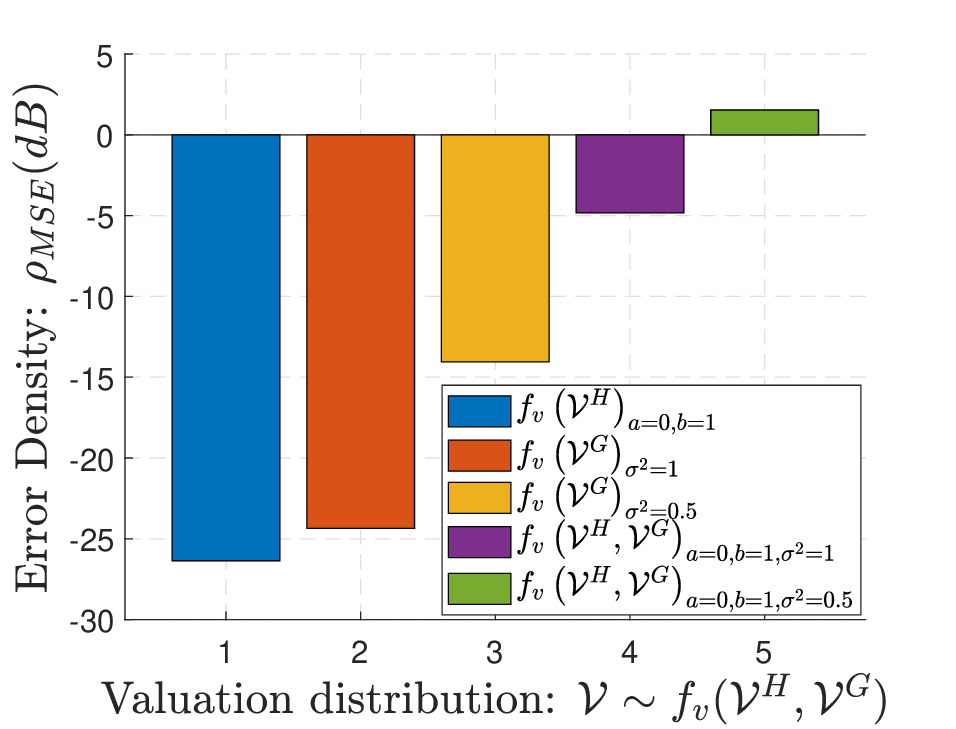}\\
\mbox{({\textit{a}})} & \mbox{({\textit{b}})} & \mbox{({\textit{c}})}\\
\end{array}$ 
\vspace{-2mm}
\caption{({\textit{a}}), ({\textit{b}})Variation of optimal BNE bidding strategy with valuation vector $\left(\mathcal{B}^{BNE}\text{ v/s }\mathcal{V}\right)$, ({\textit{c}})Variation of logarithmic mean absolute error density $\left(\rho_{MAE}(dB)~\text{ v/s }~\mathcal{V}_{i}\right)$ between analytical: \eqref{Eq: BNE_FPSB_Vh_Vg} and numerical: algorithm~\ref{Alg: numerical_fpsb} in FPSB auctions with distinct valuation distributions corresponding to $\mathcal{V}=\alpha \mathcal{V}^{H} + \beta \mathcal{V}^{G}$ and $\alpha=1$, $\beta=1$, $K=5, N=1$.}
\label{Fig: pG_fH_R_S_N} 
\vspace{-6mm}
\end{figure*}
The overall valuation metric $v_{k}$ of node $k$ can be found using the mapping function $\mathbf{f}_{v}$ as $v_{k} = \mathbf{f}_{v}(v^{1}, \dots, v^{l},\dots, v^{L})$ for the $ L$-dimensional valuation types. Here $\underline{v}^{l}$ is the lower bound of the valuation metric of each IoT node corresponding to the $l^{th}$ dimension. Ultimately, the prerequisites of \eqref{Eq: CDF_valuation_distribution} is inserted to \eqref{Eq: BNE_FPSB_general}, then the BNE of the FPSB auction for given two-dimensional valuation types, along with channel state information $\mathcal{V}^{G}$ and individual hypothesis $\mathcal{V}^{H}$ is derived as in \eqref{Eq: BNE_FPSB_Vh_Vg}. \figurename~\ref{Fig: pG_fH_R_S_N}$(a)$ shows the optimal bids $(b_{k})$ or payment strategies $(b_{k} \rightarrow q^{[n]})$ with respect to individual valuation vectors $(v_{k} \in \mathcal{V})$ from each node $k$ towards desired STFS segment $n$ for $I_{auc}=2000$ SPSB and FPSB auctions. Although the optimal BNEs are scattered with a higher variance in the SPSB auction, FPSB follows the derived analytical formula \eqref{Eq: BNE_FPSB_Vh_Vg} consistently. In addition, it is noticeable that the average of the distribution of $2^{nd}$ order statistic $\bar{\mathcal{B}}_{(2)}=N^{-1}\sum_{n=1}^{N}\left(b_{(2),n}| v\right)$, (as explained in definition~\ref{Defn: order_stat}, \cite{Order_statistics}) of the BNE scatters in the SPSB auction, aligns with the optimal strategies of FPSB auction. Consequently, an analogous relation exists between the average space of $2^{nd}$ order statistics in optimal SPSB bidding vectors and FPSB BNEs.

\vspace{-2mm}
\newtheorem{definition}{Definition}
\begin{definition} \label{Defn: order_stat}
The $p^{th}\in \{ 1, \dots, P\}$-order statistic, denoted $X_{(p)}$ is the $p^{th}$ largest realization, among $N$ draws $(N>P)$, of a random variable $X$.\footnote{The largest order statistic i.e. $X_{(p)}~=~\max\{ X_{1}, \dots, X_{p} \}$, for given random variables $X_{1}, \dots, X_{p}$ and all $p^{th}$-order statistics $X_{(p)}$, are random variables, ex: $X_{(1)}$ is the maximum of $P$ draws, $X_{(2)}$ is the second highest of $P$ draws and $X_{(P)}$ is the minimum.}
\end{definition}
In the given definition, the $ p^{th} $ order statistic, denoted as $ X_{(p)} $, refers to the position of a value within a sorted list of realizations from a random variable $ X $. Specifically, it represents the $ p^{th} $ largest value in a sequence of $ N $ independent observations of $ X $, where $ N $ is greater than $ P $, and $ p $ is an integer within the set $ \{1, \dots, P\} $. For example, if $ P $ values are drawn from the random variable $ X $, the largest value among these $ P $ draws is designated as $ X_{(1)} $, the second largest as $ X_{(2)} $, and so forth, down to $ X_{(P)} $, which is the smallest. These order statistics are themselves random variables, as they derive from the random draws of $ X $ and their values depend on the specific outcomes of these draws. This concept is fundamental in statistics for describing the distribution and properties of samples from a population.

\vspace{-2mm}
\subsection{Revenue Equivalence Theorem (RET)}
Building on the concept of order statistics as defined earlier, where \( X_{(p)} \) signifies the \( p^{th} \) largest value among a set of realizations, we can further explore its implications in auction theories, particularly in FPSB and SPSB auctions. The relevance of this statistical measure becomes clear in the Revenue Equivalence Theorem (RET), articulated in lemma \ref{Thm: RET}, known as RET, (proof: P478-480, corollary $12.24$ in \cite{maschler_solan_zamir_2013}), underpins the expected revenue outcomes in traditional auction settings.
\begin{lemma} (RET) \label{Thm: RET}
Let the bidding function $\mathbf{b}$, be a symmetric and monotonically increasing equilibrium in a sealed-bid symmetric auction with independent private values satisfying the properties of the winner node is the buyer node with the highest private value, and the expected payment of the buyer with private value 0 is 0. Then the IoT gateway receives the same expected revenue for each auction mechanism.
\end{lemma}
This lemma connects the statistical behavior of order statistics in sampling to the strategic outcomes in auction designs, demonstrating that despite differences in auction formats, the overall financial outcome for the seller (or in this case, the IoT gateway) can remain consistent, provided the bidding strategies and value distributions among participants hold certain symmetrical properties. While the BNE of SPSBs is easily derived, as shown in Lemma \ref{Thm: BNE_SPSB}, finding the BNE of FPSBs is computationally challenging due to the complexity of the derived formula \eqref{Eq: BNE_FPSB_Vh_Vg}, particularly for multi-dimensional valuations (\( >2 \)) in the generalized form \eqref{Eq: BNE_FPSB_general}. As outlined in Algorithm \ref{Alg: numerical_fpsb}, each node computes the expectation over a truncated distribution where all opponent valuations are lower. The numerical approach is conceptually simple, with RET enabling a smooth transition from trivial SPSB solutions to numerical FPSB BNEs. This is validated in \figurename~\ref{Fig: pG_fH_R_S_N}(b), which shows near-perfect alignment between analytical and numerical results, with MSE $\sim$ -5 dB for the 4th case \( f_v(\mathcal{V}^H, \mathcal{V}^G)_{a=0,b=1,\sigma^2=1} \), as in \figurename~\ref{Fig: pG_fH_R_S_N}(c), offering graphical proof of RET. However, both FPSB and SPSB employ non-cooperative, single-resource allocation strategies. Consequently, nodes with strong CSI or hypothesis valuations may exhibit selfish resource-seeking behavior, limiting interaction and hindering surplus maximization.
\vspace{-1mm}
\begin{algorithm}[t]
    \caption{Numerical Approach to Find BNE Bidding in FPSB auction}
    \label{Alg: numerical_fpsb}
    \begin{algorithmic}[1]
        \State Initialize $\mathcal{K}, \mathcal{N}, (\mathcal{V}, \mathcal{F})^{K \cross I_{auc}}$ for auction model \eqref{Eq: auc_game_model}.
        \State Calculate number of opponent players: $K^{'}=n(\{\mathcal{K} \setminus k \})$
        \State Make the truncated space w.r.t. valuation $v_{k}$: $\mathcal{V}\leq v_{k}$
        \State Drop the extra elements s.t. $\mathcal{V}_{trunc}^{K^{'} \cross \bar{I}_{auc}}; \quad \bar{I}_{auc} \in \mathbb{Z}^{+}$
        \State Reshape the valuation vector, $\mathcal{V}_{trunc}^{K^{'} \cross I_{auc}^{'}}; \quad I_{auc}^{'} \leq I_{auc}$
        \State Find winning bids in each auctioning iteration $i\in I_{auc}^{'}$
        \State Calculate the average bid: $F_{Y}(v) \cross \mathbb{E}\left[Y|Y \leq v\right]$, where $Y = \max\{\mathcal{V} \setminus v_{k}\}$.
    \end{algorithmic}
\end{algorithm}
\vspace{-4mm}
\subsection{Vickery-Clarke-Groves (VCG) Auction}
\vspace{-1mm}
Non-cooperative FPSB and SPSB auctions operate by reducing the surplus of nodes while the IoT gateway receives a higher revenue amount which is called revenue maximization (definition~\ref{Defn: revenue_maxn_min_maxminfair}$(a)$). Therefore, this phenomenon may bias the IoT gateway towards selfish gratification, enhancing the self-channel throughput for receiving signals from IoT devices, perhaps exceeding the required threshold bounds. Although data rate maximization is beneficial for the IoT gateway in terms of better QoS, it comes at a cost to IoT nodes, which consume more transmit power in the uplink than the necessary amount.
\begin{definition} \label{Defn: revenue_maxn_min_maxminfair}
A quasilinear mechanism is a revenue (a)maximizing, (b)minimizing, (c)max-min fair when, among the set of functions $c_{\mathcal{K}}^\mathcal{N}$ and $\mathcal{Q}$ that satisfy the other constraints, the mechanism selects the $c_{\mathcal{K}}^\mathcal{N}$ and $\mathcal{Q}$ that (a)maximize $\mathbb{E}_{c_{\mathcal{K}}^\mathcal{N}}\left[ \sum_{\forall k} \mathcal{Q}_{k}(b(v)) \right]$, (b)minimize $\max_{c_{\mathcal{K}}^\mathcal{N}}\left[ \sum_{\forall k} \mathcal{Q}_{k}(b(v)) \right]$, (c)maximize $\mathbb{E}_{c_{\mathcal{K}}^\mathcal{N}} \left[ \min_{k \in \mathcal{K}} v_{k} \left( c_{\mathcal{K}}^\mathcal{N} (b(v)) \right) -  \mathcal{Q}_{k}(b(v))\right]$, where $b(v)$ denotes the node's equilibrium strategy profile, \cite{multiagent_shoham_GT}.
\end{definition}
Conversely, the IoT gateway might be inclined to regulate the channel throughput threshold towards a lower stage while accepting lesser RSSIs for uplinks. Although IoT nodes are able to reduce the transmit power extensively under the revenue minimization (definition \ref{Defn: revenue_maxn_min_maxminfair}$(b)$), the gateway might have to pay the counterpart to enhance the receiver antenna gain vector for more effort to sense fewer RSSIs. Hence, the auctioneer gateway might be concerned with selecting a fair design that will be beneficial for all network entities. As a result, it is desirable to propose policies for optimal resource allocation that ensure the tricky max-min fairness conditions which describe the fairest utility as the one that makes the least-happy IoT entity the happiest as in the definition \ref{Defn: revenue_maxn_min_maxminfair}$(c)$. VCG mechanism \eqref{Eq: VCG_mechanism} which originated from the SPSB or Vickrey auction, is elegant for capturing interactive involvement among IoT uplinks and ensuring max-min fairness in simultaneous resource allocation.
\begin{subequations}\label{Eq: VCG_mechanism}
    \begin{IEEEeqnarray}{cl}
            \label{Eq: Grove_mechanism_1}
            c_{\mathcal{K}}^\mathcal{N}\left(\hat{v}\right) &= \max_{b} \sum_{k} \hat{v}_{k}(b)\\ 
            \label{Eq: Grove_mechanism_2}
            Q_{k}\left(\hat{v}\right) &= d_{k}\left(\hat{v}_{-k}\right) - \sum_{j \in \mathcal{K}\setminus k} \hat{v}_{j} \left( c_{\mathcal{K}}^\mathcal{N} \left(\hat{v}\right) \right)\\
            \label{Eq: Clarke_taxation}
            d_{k}\left(\hat{v}_{-k}\right) &= \sum_{j \in \mathcal{K}\setminus k} \hat{v}_{j} \left( c_{\mathcal{K}}^\mathcal{N} \left(\hat{v}_{-k}\right) \right)
            \vspace{-1mm}
    \end{IEEEeqnarray}
\end{subequations}
The Grove mechanism \eqref{Eq: Grove_mechanism_1}, \eqref{Eq: Grove_mechanism_2} optimizes the choice of each IoT node $c_{k}^{n}$, based on the assumption of all nodes disclosed the true utility function (P289, Theorem $10.4.2$ in \cite{multiagent_shoham_GT}) similar to in SPSB, lemma \ref{Thm: BNE_SPSB}. Under Groves's mechanism, the earning surplus of each node does not rely only on the individual resource allocation, due to imposed payments, which is explained in \eqref{Eq: Grove_mechanism_2}. Since IoT node $k$ is paid the surplus of all the other nodes under the chosen allocation, the $k$th node will be motivated towards surplus maximization of neighboring devices. 

Clarke's taxation \( d_k(\hat{v}_{-k}) \) in \eqref{Eq: Clarke_taxation} defines the payment of node \( k \), independent of its own declaration \( c_k^n \) for the STFS slot \( n \), and is based on the sum of opponent nodes’ declared valuations under the mechanism's choice. It appears as the first summation in the payment rule \eqref{Eq: Grove_mechanism_2}, while the second summation reflects the total valuation of all IoT devices except node \( k \) for the mechanism’s actual allocation. Thus, each IoT device bears its own social cost, quantifying its participation’s impact on opponents in the allocation process. If node \( k \) does not influence the mechanism's allocation, i.e., \( c_{\mathcal{K}}^\mathcal{N}(v) = c_{\mathcal{K}}^\mathcal{N}(v_{-k}) \), then the two summations in VCG payment rule \eqref{Eq: Grove_mechanism_2} cancel out, resulting in zero payment. Conversely, a nonzero payment implies node \( k \) is pivotal, meaning its presence alters the mechanism’s choice. Hence, the VCG mechanism is termed a pivot mechanism: only pivotal nodes pay. In some cases, nodes may improve opponents' allocations by participating, leading to negative social cost i.e., they are compensated by the mechanism and vice versa. The algorithm \ref{Alg: VCG_mechanism} explains the application of the VCG mechanism to find the social cost and utilities of each IoT entity by considering the assignment procedure as a knapsack problem utilizing the Hungarian algorithm, \cite{Hungarian_alg}.
\begin{algorithm}[t]
    \caption{VCG mechanism \eqref{Eq: VCG_mechanism} then, utility calculation for a specific winner node $k^{*}\in \mathcal{K}^{*}$}
    \label{Alg: VCG_mechanism}
    \begin{algorithmic}[1]
        \State Initialize $\mathcal{K}, \mathcal{N}, (\mathcal{V}, \mathcal{F})^{K \cross N}$ for auction model \eqref{Eq: auc_game_model}.
        \State Find the best assignment matrix $c_{\mathcal{K}}^\mathcal{N}$ in \eqref{Eq: Grove_mechanism_1} via Hungarian algorithm.
        \State Extract optimal valuation vector $\mathcal{V}\left( c_{\mathcal{K}^{*}}^\mathcal{N} \right)$ for winner space $\mathcal{K}^{*}$ from the valuation space $\mathcal{V}$.
        \State Calculate the total valuation for the IoT gateway except the winner node $k^{*} \in \mathcal{K}^{*}$, (second term in \eqref{Eq: Grove_mechanism_2}).
        \State Clarke taxation: Redo the Hungarian algorithm to find the best assignment for valuation space $\mathcal{V}\left( c_{\mathcal{K}^{*} \setminus k^{*}}^\mathcal{N} \right)$ except the winning node $k^{*} \in \mathcal{K}^{*}$ and take the summation, \eqref{Eq: Clarke_taxation}.
        \State Calculate the social cost $Q_{k^{*}}\left(\hat{v}\right)$ in \eqref{Eq: Grove_mechanism_2} for winner node $k^{*}\in \mathcal{K}^{*}$.
        \State Obtain the utility \eqref{Eq: utility_S_R} for the winner entity $k^{*}\in \mathcal{K}^{*}$.
    \end{algorithmic}
\end{algorithm}

While the VCG mechanism enables efficient and simultaneous STFS resource allocation, it requires IoT nodes to disclose their full valuation to the gateway. Truthful bidding can lead to overuse of resources and surplus deviation for nodes. As a gateway-centralized approach, VCG grants greater control to the gateway but limits nodes’ ability to adjust thresholds in critical phases. Acting as the mechanism designer, the gateway can impose constraints ranging from revenue maximization to max-min fairness, creating a monopolistic system with limited node autonomy. This centralized model may be unsuitable for emerging resource allocation mechanisms, particularly in edge computing scenarios where vendor-specific, constraint-independent devices demand distributed coordination. Fairness requires assessing each agent’s performance and assigning computational complexity based on the capabilities of the IoT devices. Moreover, the VCG system relies on standard auction constraints \eqref{Eq: auc_game_model} and is sensitive to non-linear bidding. Therefore, policies that promote fair and optimal resource allocation, balancing the influence between risk-driven nodes and the gateway, are essential.
\vspace{-2mm}
\section{Proposed approach } \label{Sec: 5_proposed_approach_mSAA}
The Simultaneous Ascending Auction (SAA) generalizes the English or FPSB auction, incorporating cooperative behavior among bidders and auctioneers in multi-good allocation. We discuss valuation $\mathcal{V}$ estimation for each STFS resource segment using the optimal dispersion metric $\textbf{A}$ and analyze potential risks in the modified-SAA mechanism.
\begin{algorithm}[t]
    \caption{mSAA mechanism among IoT pool and gateway}
    \label{Alg: SAA_mechanism}
    \begin{algorithmic}[1]
        \State Initialize $\mathcal{K}, \mathcal{N}, (\mathcal{V}, \mathcal{F})^{K \cross N}$ for auction model \eqref{Eq: auc_game_model}.
        \State Gateway serves the reservation price vector $\textbf{r}_{k}$ for available STFS slots to IoT device $k \in \mathcal{K}$.
        \State Initial Dispersion vector optimization, $\textbf{A}_{k}, \textbf{A} \in \mathbb{C}, k \in \mathcal{K}$.
        \If{$\mathcal{V}_{k} \geq \textbf{r}^{N \cross 1}$} \Comment{Check kick off conditions} 
            \State Inform initial bids $\mathbf{b}_{\mathcal{K}} \in \{ 0,1 \}$ from nodes to gateway.
            \State Update the price vector of STFS segments: $\textbf{q} \gets \textbf{r}+\epsilon$.
            \State \parbox[t]{\dimexpr\linewidth-\algorithmicindent}{%
            Surplus calculation \eqref{Eq: utility_S_R} and temporary resource allocation. Assign winners $\mathcal{K}^{*}$ and losers $\mathcal{K}^{'}$, $\mathcal{K}^{*} \cup \mathcal{K}^{'} \in \mathcal{K}$.%
            }
            \State \parbox[t]{\dimexpr\linewidth-\algorithmicindent}{%
            Extract the active loser space $\bar{\mathcal{K}}^{'} \subseteq \mathcal{K}^{'}$ such as $\left( \mathcal{S}_{\bar{\mathcal{K}}^{'}} \right)_{\textbf{q} \gets \textbf{q}+\epsilon} \geq 0$, in \eqref{Eq: utility_S_R} for $\mathcal{N}$ resource space.%
            }
            \While{$\bar{\mathcal{K}}^{'} \notin \textbf{0}$ $\cup$ $i \leq I^{TH}$}
                \State \parbox[t]{\dimexpr\linewidth-\algorithmicindent}{%
                Inform bids (risk or not) $\mathbf{b}_{\bar{\mathcal{K}}^{'}}:\left[( \mathcal{V}_{\bar{\mathcal{K}}^{'}} \pm (\zeta\%) v_{\bar{\mathcal{K}}^{'}} \right] \\ \in \{ 0,1 \}$ from active loser nodes to gateway.%
                }
                \State Update the price vector: $\textbf{q}_{i} \gets \textbf{q}_{i-1}+\epsilon_{i}$
                \State \parbox[t]{\dimexpr\linewidth-\algorithmicindent}{%
                Surplus calculation \eqref{Eq: utility_S_R} and temporary resource \\ allocation. Assign winners $\mathcal{K}^{*}_{i}$ and losers $\mathcal{K}^{'}_{i}$ \\ and $\mathcal{K}^{*}_{i}~\cup~ \mathcal{K}^{'}_{i}~\in~\mathcal{K}$.%
                }\label{SAA_winner_loser_swap}
                \State \parbox[t]{\dimexpr\linewidth-\algorithmicindent}{%
                Extract the active loser space $\bar{\mathcal{K}}^{'}_{i} \subseteq \mathcal{K}^{'}_{i}$ such as, \\ $\left( \mathcal{S}_{\bar{\mathcal{K}}^{'}_{i}} \right)_{\textbf{q}_{(i+1)} \gets \textbf{q}_{i}+\epsilon_{(i+1)}} \geq 0$, in \eqref{Eq: utility_S_R}  for $\mathcal{N}_{(i+1)}$ \\ resource space.%
                }\label{Alg_line: winner_conditions}
                \State \parbox[t]{\dimexpr\linewidth-\algorithmicindent}{%
                Drop weaker nodes such as, \\ $\max_{\bar{k}^{'}_{i} \in \bar{\mathcal{K}}^{'}_{i}}\left( \mathcal{S}_{\bar{\mathcal{K}}^{'}_{i}} \right)_{\textbf{q}_{(i+1)} \gets \textbf{q}_{i}+\epsilon_{(i+1)}} < 0$ permanently.%
                }\label{Alg_line: loser_conditions}
                \State Update iteration: $i \gets i+1$
            \EndWhile \Comment{Check terminal conditions}
        \State \parbox[t]{\dimexpr\linewidth-\algorithmicindent}{%
        Update Dispersion vector corresponding to the allocated optimal STFS segments, $\textbf{A} \rightarrow \textbf{A}_{c^{\mathcal{N}}_{\mathcal{K}}}$.%
        }
        \Else
            \State \parbox[t]{\dimexpr\linewidth-\algorithmicindent}{%
            Message: \textit{``The mSAA cannot be performed due to higher reservation prices for IoT device $ \forall k \in \mathcal{K}$"}.%
            }
        \EndIf
    \end{algorithmic}
\end{algorithm}
\vspace{-4mm}

\subsection{Modified Simultaneous Ascending Auction (mSAA)}
In \figurename~\ref{Fig: System_model}, the IoT gateway first issues an initial reservation price vector \( \textbf{r}_k^{N \times 1} \) for the available STFS resources to each node \( k \in \mathcal{K} \). Subsequently, the initial valuation spaces \( (\hat{\mathcal{V}}^H, \hat{\mathcal{V}}^G) \) are estimated using individual hypothesis details \( H_{i=0,\dots,M-1} \) and the optimized dispersion matrix \( \textbf{A}_k \). Nodes then provide binary responses for each offered STFS segment, indicating acceptance of current prices for surplus maximization, as detailed in Algorithm~\ref{Alg: SAA_mechanism}. The gateway assigns STFS segments \( \mathcal{N} \) to winner nodes \( \mathcal{K}^* \), where \( |\mathcal{K}^*| = |\mathcal{N}| \), aiming to maximize revenue \( \mathcal{R}_{\mathcal{N}}^{max} \); the remaining nodes form the loser set \( \mathcal{K}' \), such that \( |\mathcal{K}^*| + |\mathcal{K}'| = K \). In each iteration \( i \), a special loser subset \( \bar{\mathcal{K}}'_{i} \subseteq \mathcal{K}' \) is identified, motivated for future bidding based on the updated price vector \( \textbf{q}_{i^{+}} \). As described in Algorithm~\ref{Alg: SAA_mechanism}, line~\ref{SAA_winner_loser_swap}, a previously active loser node \( \bar{k}'_{i-1} \in \bar{\mathcal{K}}'_{i-1} \), capable of winning STFS segment \( n \in \mathcal{N} \), may enter the winner pool as \( k^*_{i} \in \mathcal{K}^*_{i} \). 

Simultaneously, a winner node \( k^*_{i-1} \in \mathcal{K}^*_{i-1} \) with no further bidding intent for the same segment is demoted to the loser pool \( k'_{i} \in \mathcal{K}'_{i} \). This node may still compete for another STFS slot or the same slot in future iterations \( i^{+} \). Thus, nodes are inter-swapped between winner and loser pools toward a global equilibrium in a converging manner. The potential for further bidding is defined by the non-negativity of instantaneous surplus \( \mathcal{S}_{\bar{k}'_{i}} \) in \( \bar{\mathcal{K}}'_{i} \). Nodes violating marginal conditions in iteration \( i \) are permanently excluded (see lines~\ref{Alg_line: winner_conditions} and~\ref{Alg_line: loser_conditions}). The auction ends if no active losers remain or if the iteration limit \( I^{TH} \) is reached. Finally, dispersion vectors are adjusted along an orthogonal frame to minimize uplink interference and ensure RSSI satisfaction, enhancing gateway signal throughput.

\begin{figure}[!t]
\centering
$\begin{array}{cc}
\includegraphics[width=0.48\columnwidth, trim={0.5mm 2.5mm 9mm 0mm},clip]{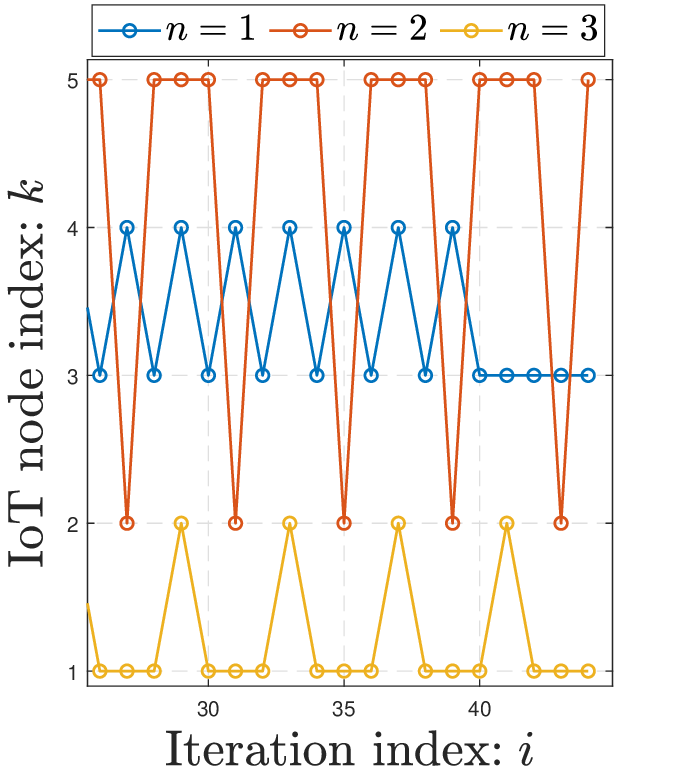}
\label{Fig: SAA_node_interaction_plot} &
\includegraphics[width=0.40\columnwidth, trim={0mm 0mm 0mm 0mm},clip]{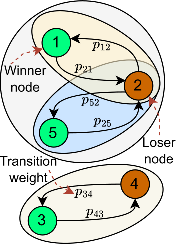}
\label{Fig: SAA_node_interaction_state}\\
\mbox{({\textit{a}}) Convergence patterns} & \mbox{\vspace{-5mm}({\textit{b}}) State diagram}\\
\end{array}$ 
\vspace{-1mm}
\caption{$(a)$~Convergence patterns, and $(b)$ State graph  of indicator metric $c_{\mathcal{K}}^{\mathcal{N}}$ for $K~=~5$ IoT nodes auctioning $N=3$ STFS resource segments over $I=44$ processing iterations.}
\label{Fig: SAA_node_interaction_both}
\vspace{-8mm}
\end{figure}

\figurename~\ref{Fig: SAA_node_interaction_both}$(a)$ illustrates IoT bidders' internal behaviors in the mSAA mechanism, each competing for an optimal STFS slot across random iteration blocks until global allocation is achieved. The diagram shows instantaneous hopping patterns as nodes demand resources: the 1st node secures the 3rd STFS slot in 3 iterations; the 2nd node alternates bids for the same slot and also tests suitability for the 2nd slot, entering an auction with node 5. Similarly, nodes 3 and 4 bid for the 1st slot, with node 3 winning. These interactions form clusters, shown in \figurename~\ref{Fig: SAA_node_interaction_both}$(b)$, exhibiting oscillatory convergence through binary voting with weighted probabilities \( p_{k^*_i, \bar{k}'_i} \) sent to the gateway. This interaction reveals correlated bidding behaviors and prevents monopolistic strategies via distributive sub-auctioning. The decentralized architecture enables nodes to self-identify compatible STFS segments using simple, low-hypothesis decisions in heterogeneous networks. Thus, mSAA promotes fairer surplus distribution, reduces transmission power, and prioritizes resource allocation efficiently. Additionally, nodes adaptively make decisions through periodic optimal journeys, navigating dynamic \( M \)-hypotheses and predicting future margins using recent memory blueprints—minimizing both randomness and iteration count.

\emph{Properties}: Three main properties can be guaranteed in the proposed mSAA mechanism in algorithm \ref{Alg: SAA_mechanism},
\begin{itemize}
    \item Incentive Compatibility (truthfulness): In the standard SAA, the bidders earn the maximum surplus gain by submitting binary bidding responses truthfully. For instance, let $b_{k}^{n}[v_{k}]$ denote the binary bidding response of bidder $k$ based on true valuation $v_{k}$ against STFS $n^{th}$ resource block, $(v_{k} - q_{k}^{n})$ is the maximum surplus utility. However, the mSAA mechanism has the flexibility to deviate from the true valuation and do underbidding or overbidding while regulating the stable state of the allocation mechanism. Although underbidding violates or weaker the incentive capability in mSAA, it is robust with average surplus degradation of the entire auctioning result as explained in the results section \ref{Sec: 6_results_discussion}. 
    \item Individual Rationality: The payment of bidder $k$ should be less than the true valuation itself for non-negative utilities. The algorithm \ref{Alg: SAA_mechanism} (line \ref{Alg_line: loser_conditions}), eliminates the losers who are violating the property of individual rationality and extracts active loser space $\bar{\mathcal{K}}^{'}, \quad \forall k \in \mathcal{K}$.
    \item Price monotonicity: The price $q_{k}^{n}$ of each STFS resource block $n \in \mathcal{N}$ for each IoT bidder $k \in \mathcal{K}$ is increasing monotonically by step size $\epsilon$ in each iteration. Then, given the set of bidding except the winning bid $b[q_{k}^{n}]_{i}$ in $i^{th}$ iteration, then any bid $b[q_{k}^{n}]_{i+}$ also wins with $[q_{k}^{n}]_{i+} \geq [q_{k}^{n}]_{i}$ in the following iterations.    
\end{itemize}

In extremely dense IoT regimes with $K > 100$ devices, the iterative signaling in mSAA leads to non-negligible control-channel utilization. Each iteration requires the gateway to broadcast updated reservation prices while active losers submit binary bids, resulting in a measured peak overhead of approximately $37\%$ of the available control bandwidth. This reflects the high contention level in large-scale networks. To mitigate the risk of congestion, mSAA employs several safeguards: (i) a hard iteration cap $I_{\mathrm{TH}}$ that bounds the maximum number of updates, (ii) binary signaling rather than high-dimensional valuation reports, (iii) pruning to an active-loser set $\bar{\mathcal{K}}'$ so that only bidders with nonnegative marginal surplus remain eligible, and (iv) the option of batching or piggybacking price updates onto existing control frames. These design choices guarantee predictable convergence and prevent control-channel saturation even in dense IoT deployments.

By contrast, the conventional VCG mechanism lacks an inherent iteration definition or a maximum iteration safeguard. In practice, the absence of a bounded stopping criterion can lead to oscillatory or divergent behavior in dense networks, particularly when misreporting occurs. For fairness in our experiments, we introduced an external maximum recomputation budget $I_{\mathrm{VCG}}^{\max}$ together with a tolerance-based stopping rule; however, we emphasize that these modifications are not part of classical VCG and highlight its practical vulnerability in large-scale IoT environments. The explicit convergence safeguards in mSAA therefore provide a distinct robustness advantage compared with VCG.

\emph{Complexity}: Traditional combinatorial auction-based resource allocation requires a centralized exhaustive search over an NP-hard space, with complexity \( \mathcal{O}(K^N) \) for \( K \) IoT bidders and \( N \) STFS resource blocks. In contrast, the proposed mSAA algorithm distributes computation among IoT entities. Each bidder \( k \) provides binary responses over all possible STFS block combinations, totaling \( \sum_{n=1}^{N} \binom{N}{n} = 2^N - 1 \) possibilities per iteration.
Thus, the overall complexity becomes  $\mathcal{O}\!\big(\bar{K}'\cdot(2^N-1)+I\big)$, where $\bar{K}'\le N \ll K$ maps to the number of resource gaps. With a small, system-selected $N$ and a hard iteration cap $I_{\mathrm{TH}}$, runtime is bounded even in ultra-dense regimes.
And \( I \) is the total number of iterations with price updates \( q_{(i)} \gets r + \epsilon \cdot i \). IoT entities may be allocated suboptimal resources during the intermediate stages of convergence through $I$ iterations toward global equilibrium. In the worst case, with \( I = (q_{(i)}^{\text{max}} - r)/\epsilon \), mSAA’s complexity remains significantly lower than \( \mathcal{O}(K^N) \), i.e., $\mathcal{O}\left(\bar{K}' \cdot (2^N - 1) + \frac{q_{(i)}^{\text{max}} - r}{\epsilon}\right) < \mathcal{O}(K^N), \quad \bar{K}' \leq K.$ Furthermore, if each active loser bids for only one STFS block, complexity reduces further to $\mathcal{O}\left(\bar{K}' \cdot N + \frac{q_{(i)}^{\text{max}} - r}{\epsilon}\right).$
\vspace{-2mm}
\subsection{Dispersion Matrix Optimization} \label{sec_sub: dispersion_mat_opt_A}
After the auction concludes, IoT nodes receive their allocated STFS blocks \( c_{\mathcal{K}}^{\mathcal{N}} \), distributed orthogonally across space, time, and frequency to boost channel throughput \eqref{Eq: throughput_x}. Despite the auction enabling independent transmissions, slight deviations may occur due to time delays and frequency offsets from channel and hardware impairments. To address this, each node \( k \in \mathcal{K} \) adjusts its uplink gain and phase to meet RSSI limits and reduce interference at the gateway. This leads to the formulation of a dispersion metric optimization problem \( \textbf{a}_{k} \in \mathbf{A} \) for node \( k \), accounting for interference from other nodes \( j \in \mathcal{K} \setminus k \).
\begin{subequations}\label{Eq: A_opt_dispersion}
\begin{IEEEeqnarray}{cl}
\underset{\textbf{a}_{k} \in \mathbf{A}}{\mbox{minimize}}&~~ \sum_{k =1}^{K} \Vert\textbf{z}_{k} - \textbf{z}_{k}^{*}\Vert^2 \label{Eq: A_opt_obj_fn}
\\ \mbox{subject to}&~~ \textbf{z}_{k} = \textbf{a}_{k}\textbf{g}_{k}\textbf{x}_{k} + \textbf{w}_{k}, \quad \textbf{z}_{k}^{*} = c_{k}^{[n]}s_{k}, \eqref{Eq: tx_signal} \label{Eq: A_opt_z}
\\&~~ -\pi \leq \angle{\textbf{a}_{k}} \leq \pi \label{Eq: A_opt_phase_bounds}
\\&~~ \eqref{Eq: Tx_signal_STFS}, \eqref{Eq: tx_signal_power_avg}, \eqref{Eq: throughput_x}, \eqref{Eq: sub_opt_revenue}, \quad \forall k\in \mathcal{K} \label{Eq: A_opt_rest_eqn} 
\end{IEEEeqnarray}
\end{subequations}
In \eqref{Eq: A_opt_dispersion}, each IoT node aligns its uplink vector \( \textbf{z}_k \) with the allocated resource vector \( \textbf{z}_k^* \) as defined in \eqref{Eq: A_opt_z}, minimizing the Euclidean distance deviation across all devices. The STFS elements are i.i.d. and structured orthogonally, guided by the gateway through the allocation indicator \( c_{\mathcal{K}}^{\mathcal{N}} \). Uplink distortions from the reciprocal channel \( \textbf{g}_k \) and hardware noise \( \textbf{w}_k \) are corrected by adjusting the dispersion phase jitters \eqref{Eq: A_opt_phase_bounds}, while signal attenuation is managed via gain control, constrained by transmit power bounds \eqref{Eq: Tx_signal_STFS}, \eqref{Eq: tx_signal_power_avg} in \eqref{Eq: A_opt_rest_eqn}. 

\begin{figure}[!t]
\centering
\includegraphics[width=0.8\columnwidth, trim={3mm 4mm 11mm 14mm},clip]{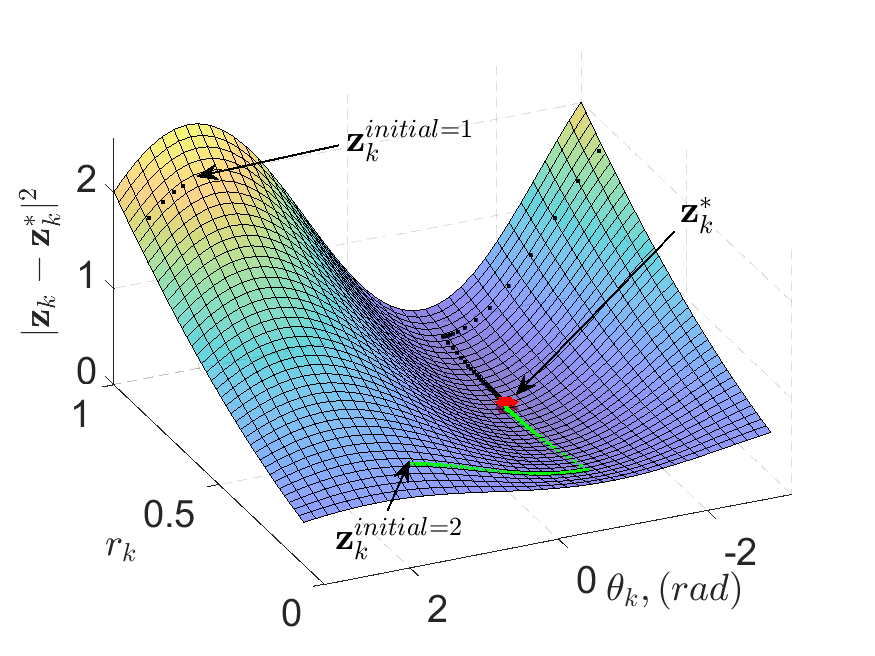}
\caption{The convergence pattern from a random initial point $\textbf{z}_{k}$, toward the global stationery $\textbf{z}_{k}^{*}$, along the dimensions of phase and gain while occurring optimal dispersion vector $\textbf{a}_{k}$ for each IoT device $k \in \mathcal{K}$.}
\vspace{-3mm}
\label{Fig: r_theta_opt_A}
\end{figure}
\figurename~\ref{Fig: r_theta_opt_A} illustrates the non-convex landscape of the dispersion optimization objective in low-gain regions; nevertheless, the objective in~\eqref{Eq: A_opt_obj_fn}, written explicitly as \(\sum_{k \in \mathcal{K}} \!\left(r_k^2 + {r_k^*}^2 - 2\,r_k\,r_k^* \cos(\theta_k - \theta_k^*)\right)\), is \emph{periodic in the phase domain} by virtue of Euler’s representation \(\mathbf{z}_k = r_k e^{j\theta_k}\). This periodicity enables a convexification-by-separation strategy in which the optimization is decomposed into two stages: first, for fixed gain \(r_k>0\), we optimize the phase \(\theta_k\) so as to minimize the phase misalignment \(\theta_k - \theta_k^*\) (a convex problem in the wrapped angle); second, holding the resulting phase \(\theta_k^*\) fixed, we adjust the gain to align the amplitude \(r_k\) with the target \(r_k^*\). Starting from an arbitrary initial condition determined by instantaneous channel and hardware imperfections, each IoT node converges to the global optimum corresponding to its allocated STFS block \(c_k^{[n]}\); in practice, low-complexity nodes implement this with simple gradient descent, and the minimal corrective action is captured by the optimized dispersion element \(a_k^*\). 

The first (phase) stage rotates each uplink phasor toward its reference orientation \(\theta_k^*\), thereby coherently aligning signals with the orthogonal STFS frame and reducing inter-stream interference; the second (gain) stage then scales the phasor to the required magnitude \(r_k^*\), meeting RSSI and throughput targets without excess power. Because phase errors dominate interference coupling, correcting \(\theta_k\) first yields a well-behaved (effectively convex) descent direction even when \(r_k\) is small, after which amplitude adjustment becomes a one-dimensional refinement. Consequently, the two-stage procedure steers transmissions onto energetically efficient operating points: interference is suppressed, required transmit power is lowered, and the gateway observes cleaner, near-orthogonal superpositions consistent with the allocation indicated by \(c_k^{[n]}\).

\begin{figure}[!t]
\centering
\includegraphics[width=0.8\columnwidth, trim={1mm 20mm 13mm 0.6mm},clip]{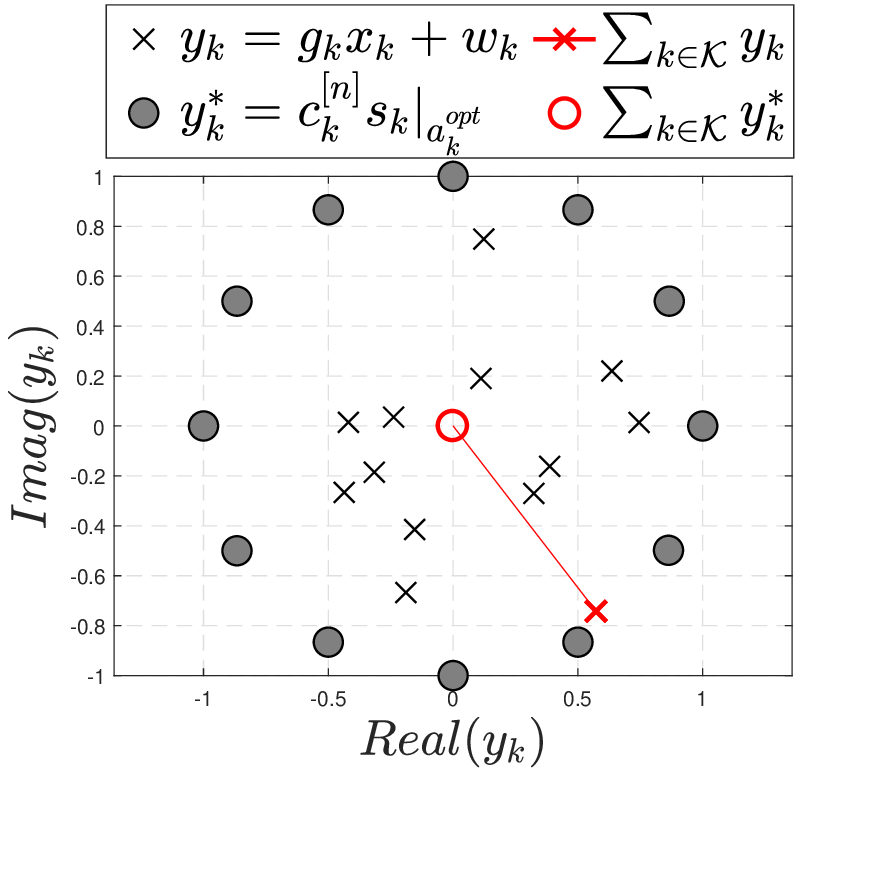}
\caption{The strength and time/frequency deviation of each uplink $k$ at the gateway with and without pre-compensation utilizing optimal dispersion metric $\textbf{A}$ for $K=12$ nodes.}
\label{Fig: dispersed_yk_s}
\vspace{-4mm}
\end{figure}
\figurename~\ref{Fig: dispersed_yk_s} illustrates the amplitude and phase of uncompensated uplinks \(y_k\) (per-device observations) together with their superposition \(\sum_{k\in\mathcal{K}} y_k\) at the gateway after propagation through a noisy medium. The aggregated waveform captures the composite interference pattern and residual time/frequency misalignments, revealing why receiver-sensitivity and target–data-rate thresholds can be difficult to satisfy for each uplink within its allocated STFS segment when dispersion is not corrected. After applying the optimized dispersion metric \(a_k^*\), each device pre-equalizes its transmission so that the effective uplink aligns with the i.i.d.\ STFS structure imposed by the auction (i.e., phase is steered toward \(\theta_k^*\) and amplitude toward \(r_k^*\) for the allocated block). As a result, inter-stream interference is attenuated, gain/phase distortions are compensated, and the received signal characteristics meet the channel–throughput requirements for reliable communication.

The uncompensated sum \(\sum_k y_k\) represents a non-coherent superposition of phasors with random rotations and unequal magnitudes; these misalignments increase cross-talk across orthogonal STFS dimensions and effectively raise the power required to achieve a given error-rate or throughput. The per-node adjustment \(a_k^*\) operates as a lightweight, transmitter-side phasor correction: first rotating each uplink toward its reference orientation and then scaling it to the desired magnitude associated with its STFS allocation. This restores near-orthogonality among streams at the gateway, lowers the interference floor, and reduces the transmit energy needed to satisfy RSSI and rate targets—thereby converting a collision-prone, power-inefficient aggregate into a structured, energy-efficient multi-access signal consistent with the allocation outcome.

The proposed dispersion optimization handles channel variability and imperfect CSI via a two-stage update per device—first aligning phase $\theta_k$ to the current estimate, then scaling gain $r_k$ to the target for its STFS segment—recomputed at each mSAA iteration using instantaneous CSI with stable first-order (gradient) updates and negligible overhead. Physically, these iterative pre-equalization steps restore near-orthogonality across streams, mitigate interference under mobility/drift, and reduce the transmit energy required to meet throughput targets in dense IoT deployments. While \eqref{Eq: signal_model} adopts a normalized additive-interference model with AWGN for tractability, this serves as a baseline abstraction to isolate the auction dynamics. More advanced PHY effects such as frequency-selective fading, correlated multi-user interference, or non-Gaussian noise map to equivalent SNR shifts or extended dispersion adjustments, and thus do not alter the qualitative conclusions regarding mSAA’s convergence and efficiency.  
%
%
%

\begin{figure*}[!t]
\centering
\includegraphics[width=\linewidth, trim={0mm 0mm 0mm 0mm},clip]{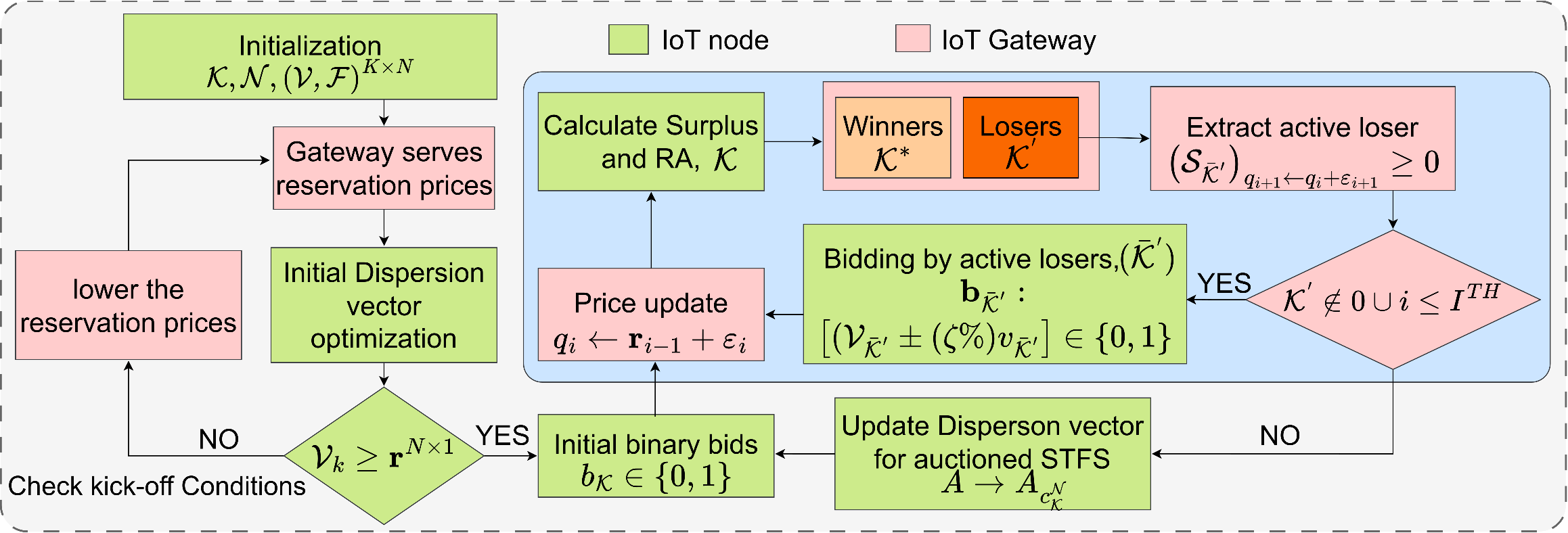}
\caption{Diagram of the mSAA mechanism mirrored to proposed Algorithm~\ref{Alg: SAA_mechanism}.}
\label{Fig: mSAA_diagram}
\end{figure*}
In order to map the mSAA mechanism into an IoT network-based simulation scenario, we have presented a flow-based representation in \figurename~\ref{Fig: mSAA_diagram} that directly mirrors the logical steps in Algorithm~3. The diagram highlights the iterative exchange between IoT devices and the gateway during the auction process. At initialization, the gateway broadcasts reservation prices for all available STFS segments and performs an initial dispersion vector optimization. IoT nodes then evaluate their valuations relative to the reservation prices and, if conditions are satisfied, submit binary bids for candidate segments. Based on these bids, the gateway updates prices monotonically in step size $\varepsilon$, calculates current surplus utilities, and temporarily allocates winners and losers. A critical feature emphasized in the diagram is the extraction of the \emph{active loser set}, i.e., those nodes that retain positive marginal surplus and are eligible to re-enter competition in subsequent iterations. Active losers may adjust their strategies under updated prices and re-bid, while nodes violating individual rationality are permanently removed. The gateway continues this iterative price update and bidding cycle until no active losers remain or the iteration cap $I_{\mathrm{TH}}$ is reached. At convergence, the dispersion vectors are finalized for the optimally allocated STFS segments, ensuring robust alignment between channel conditions and assigned resources.


\vspace{-1mm}
\section{Numerical Results and Discussion} \label{Sec: 6_results_discussion}
This section presents simulation results for optimal STFS resource assignment by the gateway, evaluating scalability based on IoT pool size and resource capacity. The proposed mSAA mechanism is compared with standard auction methods—FPSB, SPSB, and VCG.
Simulations assume unity weights \( \alpha, \beta = 1 \), equalizing the influence of channel gain and node hypothesis. 
Rayleigh variance is set to \( \sigma^2 = 1 \) without loss of generality, and the uniform distribution upper bound to \( b = 1 \), normalizing transmit power through equal-variance complex Gaussian vectors and equi-probable hypothesis, respectively. These standard statistical distributions in communication metrics are analytically tractable in the simulation results; however, the proposed mSAA algorithm is capable of dealing with empirical data acquired through real-world aspects such as non-line of sight (NLOS), asymmetric heterogeneity, and user mobility. Although the quantitative data in resource allocation vary for empirical datasets, the qualitative patterns among communication metrics would be secured. Scenarios focus on dense networks with \( K \geq N \), where the number of IoT nodes meets or exceeds available STFS slots. Scalability analysis is detailed in Section~\ref{Sec: 5_proposed_approach_mSAA}. Additionally, results highlight cases with multiple bidders showing risk-prone behaviors like underbidding or overbidding, which amplify dynamic network interactions. 
Next we outline the simulation environment, including the assumptions, default values, and variations of the key parameters. The simulation scenario consists of a single gateway coordinating $N$ orthogonal STFS segments among $K$ IoT devices, with $K \ge N$ to reflect dense IoT conditions. Each gateway–device link is modeled as flat Rayleigh fading with AWGN. The detailed parameter settings and the rationale for their selection are summarized below.

\begin{itemize}[leftmargin=*]
  \item \textbf{Number of devices ($K$):} Varied between $10$ and $40$ with $K \ge N$.  
  \emph{Rationale:} Captures dense IoT conditions where resource demand exceeds supply, highlighting the surplus–revenue trade-offs and stressing the dynamics of the mSAA allocation mechanism.
  
  \item \textbf{Number of STFS segments ($N$):} Default $N=10$; swept in $[5, 30]$ for scalability.  
  \emph{Rationale:} $N=10$ serves as a balanced baseline for revenue–surplus evaluation, while sweeping $N$ explores scalability under both resource-scarce and resource-rich regimes.
  
  \item \textbf{Channel fading variance:} Rayleigh fading with unit variance.  
  \emph{Rationale:} Provides a normalized fading model independent of absolute SNR; different propagation conditions map to SNR shifts without affecting comparative outcomes.
  
  \item \textbf{Noise:} AWGN with normalized variance.  
  \emph{Rationale:} A standard assumption to isolate allocation effects from variations in noise power.
  
  \item \textbf{Valuation weights $(\alpha,\beta)$:} Set to $(1,1)$.  
  \emph{Rationale:} Ensures that the hypothesis term ($V_H$) and channel gain term ($V_G$) contribute equally to the total valuation $V$, consistent with the theoretical framework in Section II.
  
  \item \textbf{Hypothesis prior bounds $(a,b)$:} Uniform distribution over $[0, 1]$.  
  \emph{Rationale:} Produces bounded, interpretable equiprobable priority values and ensures stable surplus and revenue computations.
  
  \item \textbf{Reservation price vector $\mathbf{r}$ and price increment $\varepsilon$:} Gateway initializes $\mathbf{r}$ and increases prices monotonically by $\varepsilon$.  
  \emph{Rationale:} Monotone increments guarantee convergence; a sufficiently small $\varepsilon$ balances stability and computational efficiency.
  
  \item \textbf{Risk/misreporting parameter ($\zeta$):} Baseline $\zeta=0\%$ (truthful bidding); robustness tested up to $\zeta=40\%$.  
  \emph{Rationale:} Models strategic misreporting (underbidding/overbidding) in heterogeneous IoT environments, enabling evaluation of robustness of the mSAA mechanism.
  
  \item \textbf{Iteration cap ($I_{\mathrm{TH}}$):} Auction stops when no active losers remain or the cap is reached.  
  \emph{Rationale:} Ensures practical runtime limits while preserving convergence properties of the iterative auction.
\end{itemize}

All simulation results are averaged over multiple independent channel and hypothesis realizations. Scalability is examined by sweeping $N$ while fixing $K$ (and vice versa), as specified in the figure captions. After allocation, dispersion vectors are optimized as described in Section~II to mitigate channel and hardware impairments prior to throughput and surplus/revenue evaluation. It is worth-mentioning here that our results explores a wide transmit-power sweep of $2$–$90$W to benchmark mSAA and exhaustive-search baselines across a broad operating region. However, we emphasize that practical IoT devices operate at $P_{\mathrm{tx}} < 10$W. The lower part of the sweep ($2$–$10$W) therefore represents the deployment-relevant regime, where mSAA retains its efficiency advantage while remaining within realistic power budgets. Results at higher power levels should be interpreted as scalability tests rather than literal IoT device operation points.

\vspace{-3mm}
\subsection{Average Productive Surplus vs Network Capacity}
\begin{figure}[t!]
\centering
\includegraphics[width=0.8\columnwidth, trim={0.0mm 2.0mm 9mm 11mm},clip]{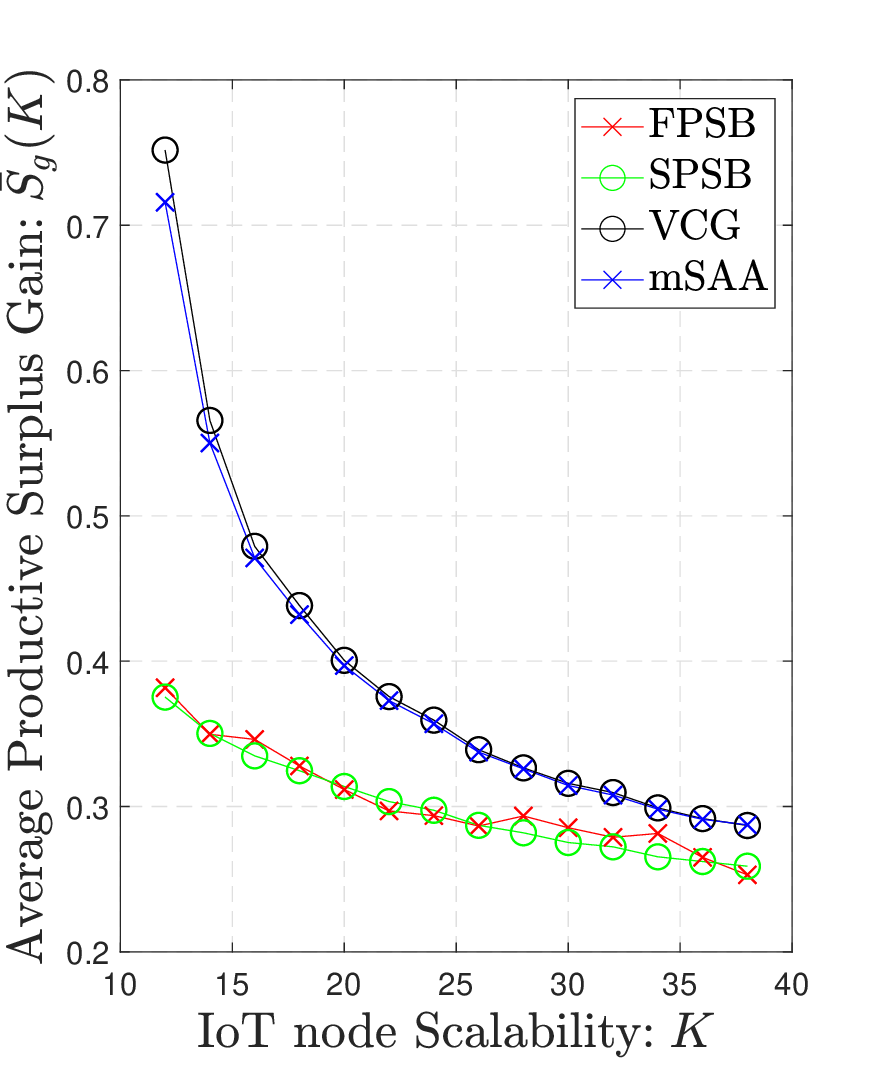}\\
\caption{Variation of normalized average surplus with number of IoT nodes ($\bar{S}_{g}$ vs $K$) in the network for $\mathcal{V}=\alpha \mathcal{V}^{H} + \beta \mathcal{V}^{G}$ and $\alpha=1$, $\beta=1$, $a=0$, $b=1$, $\sigma^{2}=1$, $N=10$ over risk-free bidding ($\zeta = 0\%$).}
\vspace{-4mm}
\label{Fig: S_K_zeta_0}
\end{figure}

\begin{figure}[t!]
\centering
\includegraphics[width=0.8\columnwidth, trim={0.0mm 2.0mm 9mm 11mm},clip]{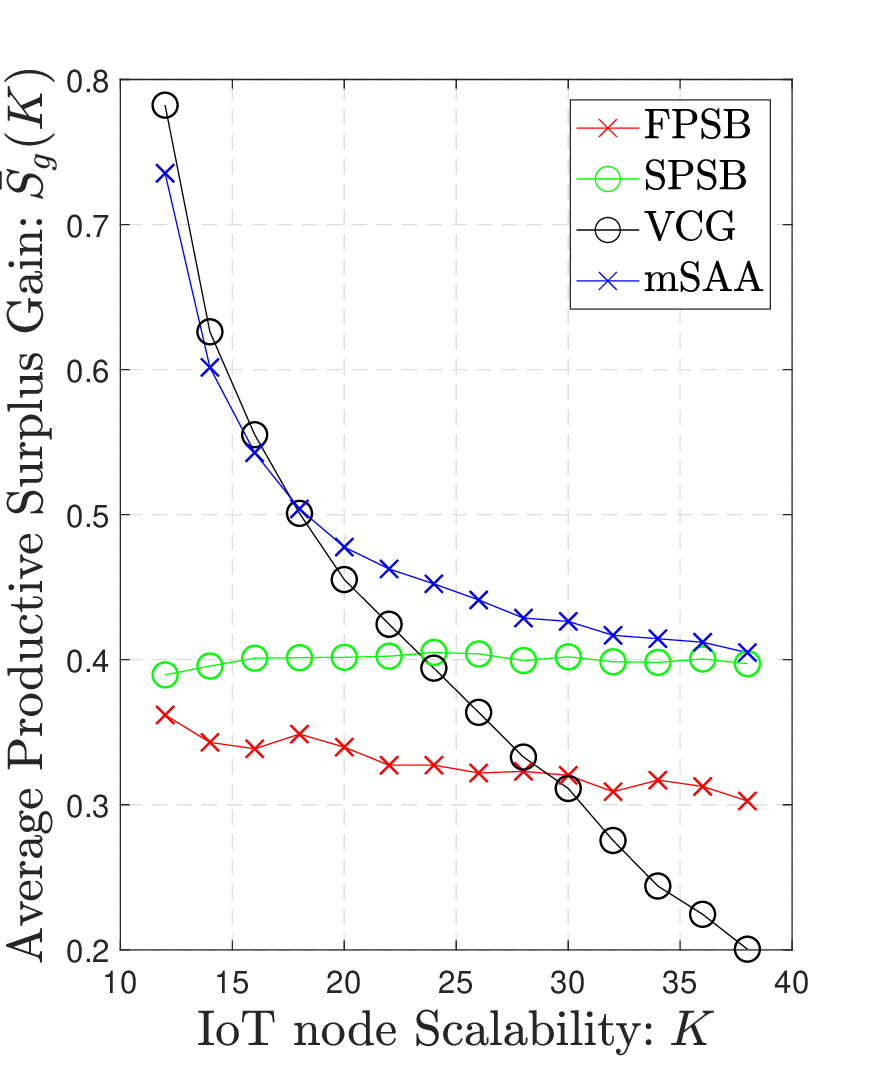}
\caption{Variation of normalized average surplus with number of IoT nodes ($\bar{S}_{g}$ vs $K$) in the network for $\mathcal{V}=\alpha \mathcal{V}^{H} + \beta \mathcal{V}^{G}$ and $\alpha=1$, $\beta=1$, $a=0$, $b=1$, $\sigma^{2}=1$, $N=10$ over underbidding ($\zeta = 4\%$).}
\label{Fig: S_K_zeta_4}
\end{figure}

\figurename~\ref{Fig: S_K_zeta_0} characterizes how competitiveness among IoT nodes intensifies as network density increases (larger $K$), which manifests as a decrease in average productive surplus per node. \textbf{A clear surplus gap ($75\% \rightarrow 16\%$ for $K = 10 \rightarrow 25$) emerges between traditional sealed-bid baselines (FPSB, SPSB) and the cooperative mechanisms (VCG, mSAA), especially in the regime where the effective IoT capacity satisfies $K' < K/2$, indicating limited scalability of classical auctions under congestion.} This gap is consistent with the strategic nature of the baselines: in FPSB/SPSB, bidders act selfishly and independently, with no coordination across users; by contrast, VCG and mSAA exploit Bayesian structure and, for mSAA in particular, iterative price adjustment to better align allocation with valuations, yielding higher surplus and lower outage probability in dense IoT conditions. Notably, FPSB and SPSB exhibit nearly overlapping surplus curves, consistent with the revenue equivalence theorem (RET; Lemma~\ref{Thm: RET}), implying similar per-node benefits. Likewise, VCG and mSAA—each derivable from classical auction principles—also align closely due to RET, while still outperforming FPSB/SPSB in both surplus gain and energy usage owing to their stronger efficiency properties.

When the auction includes a special bidder subset $\mathcal{K}_\zeta$ that deviates from truthful behavior, the evaluation framework changes; as shown in \figurename~\ref{Fig: S_K_zeta_4}, some IoT devices underbid using reduced valuations $v_k' \in \mathcal{V}_\zeta \;\leftarrow\; \{ v \in \mathcal{V} \mid v - \max(v)\cdot \zeta\% \}$, which is consistent with the formal notion that $v_k'$ is an underbid with respect to $v_k$ if $v_k'(n) \le v_k(n)$ for all $n \in \mathcal{N}$ and $v_k'(n) < v_k(n)$ for at least one $n \in \mathcal{N}$ (cf.~\cite{overbid_underbid_theory_01}); symmetrically, overbidding corresponds to $v_k'(n) \ge v_k(n)$ for all $n \in \mathcal{N}$ with strict inequality for at least one $n \in \mathcal{N}$. This risk-driven strategy aims to lower individual payments while attempting to preserve or increase the bidder’s surplus. Empirically, \figurename~\ref{Fig: S_K_zeta_4} shows that slight underbidding can mitigate the decline in average surplus under high competition: SPSB tends to expand achievable surplus vectors while FPSB slows the rate of surplus reduction, and the proposed mSAA maintains a high surplus with only a minimal drop as density grows, performing on par with SPSB in dense regimes. \textbf{Quantitatively, \figurename~\ref{Fig: S_K_zeta_4} shows that when the number of devices $K$, increases from 10 to 40 with \(N=10\), the proposed mSAA achieves approximately 33--100\% higher system surplus compared with FPSB and 0--100\% higher compared with SPSB. In contrast, VCG degrades sharply under risk (misreporting), exhibiting a rapid surplus decrease (approximately 0--100\% compared to mSAA for K=17--40) relative to the risk-free case}; this sensitivity stems from the violation of VCG’s truth-telling (dominant-strategy) property (Lemma~\ref{Thm: BNE_SPSB}), which undermines the efficiency of the social-cost computation in~\eqref{Eq: Grove_mechanism_2} and destabilizes the Clarke tax in~\eqref{Eq: Clarke_taxation}, making VCG less robust to parameter deviations and bidder collisions in heterogeneous IoT settings.

The results in \figurename~\ref{Fig: S_K_zeta_0} and \figurename~\ref{Fig: S_K_zeta_4} physically represent how resource contention in a dense IoT network translates into energy and efficiency trade-offs. When the number of devices $K$ grows relative to the number of orthogonal resources $N$, more users compete for fewer slots. This forces assignments onto links with weaker instantaneous channel gains or less favorable hypothesis valuations, meaning devices must expend more transmit power to maintain target throughput. Consequently, the average productive surplus decreases as network density rises. Mechanisms such as mSAA and VCG under truthful bidding allocate resources preferentially to devices with stronger links, thereby aligning resource usage with better SNR conditions, reducing power consumption, and sustaining higher system-wide surplus. Under underbidding ($\zeta > 0$), however, weaker devices artificially inflate their chances of winning slots, which results in allocations to energetically inefficient links. This increases the physical transmit energy required for reliable communication and degrades system surplus. mSAA mitigates this effect through its iterative price adaptation, which gradually suppresses inefficient allocations and restores efficiency, while VCG collapses under misreporting because it depends critically on truthfulness for correctness. Thus, these results represent the fundamental physical tension between truthful versus strategic reporting: truth-aligned mechanisms concentrate resources on favorable channels and minimize energy use, whereas misreporting mechanisms may divert resources to poor links, increasing system energy consumption and lowering effective surplus.

\subsection{Average Gateway Revenue vs IoT Surplus}
\begin{figure}[t!]
\centering
\includegraphics[width=0.8\columnwidth, trim={0.0mm 2.0mm 9mm 11mm},clip]{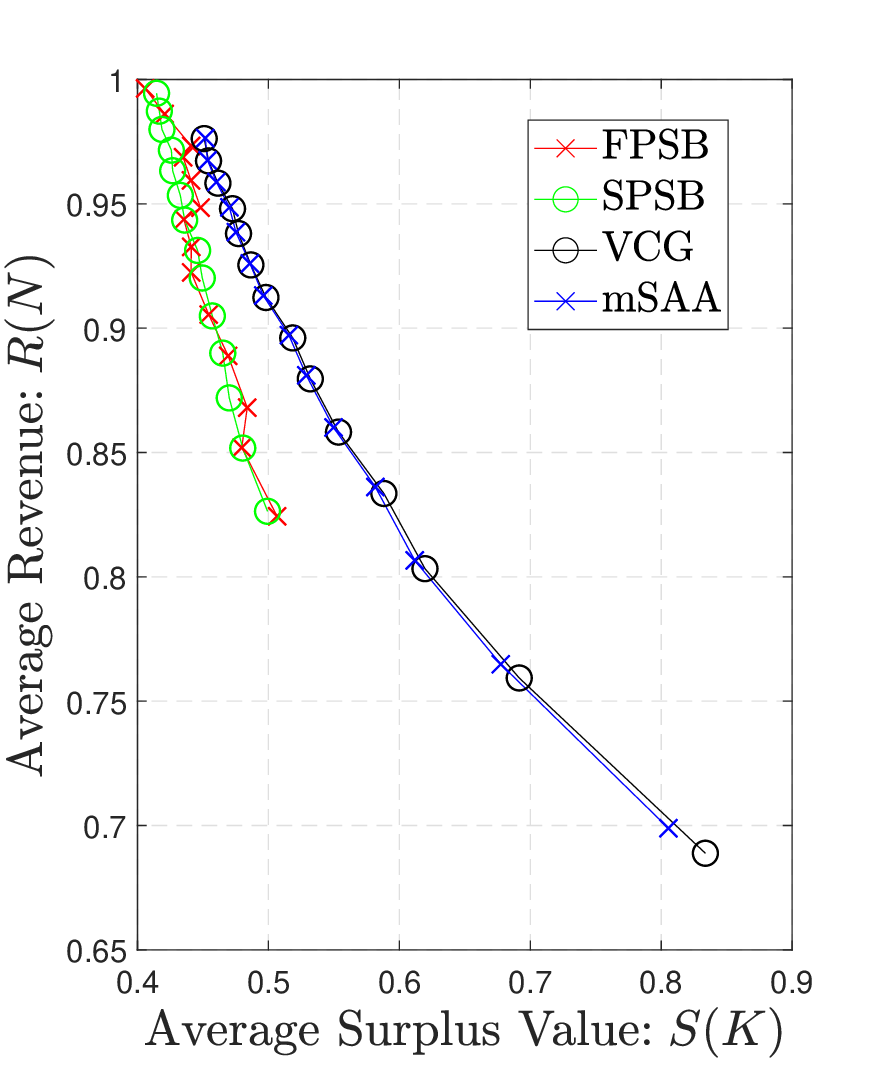}
\caption{Relation between normalized average revenue and the normalized average surplus ($\bar{R}$ vs $\bar{S}$) with the varying number of IoT nodes $K$ in the network for $\mathcal{V}=\alpha \mathcal{V}^{H} + \beta \mathcal{V}^{G}$ and $\alpha=1$, $\beta=1$, $a=0$, $b=1$, $\sigma^{2}=1$, $N=10$ over risk-free bidding ($\zeta = 0\%$).}
\label{Fig: R_S_zeta_0} 
\end{figure}
\figurename~\ref{Fig: R_S_zeta_0} depicts the joint behavior of gateway revenue and bidder surplus for a fixed resource budget of $N=10$ STFS segments while the IoT network size varies, revealing the characteristic trade-off frontier between the two metrics. An inverse relationship is evident: as competition intensifies with larger $K$, nodes expend more transmit power to remain competitive and to maintain fairness, which strengthens uplinks and lowers collision probability, thereby \emph{increasing gateway revenue} while \emph{reducing average node surplus}; when competition eases, the effect reverses. This trade-off induces an operating equilibrium in which the allocation of orthogonal STFS segments balances gateway-centric (revenue) and node-centric (surplus) performance. \textbf{In this risk-free regime, FPSB and SPSB produce similar outcomes—moderately high surpluses for nodes but slightly diminished gateway revenue—consistent with the revenue equivalence theorem (RET) behavior in single-slot classical auctions; nevertheless, the achieved node surpluses remain below half of the normalized maximum \emph{(best performance point: approximately surplus=0.5 for revenue=0.83)}, indicating limited efficiency under congestion. By contrast, VCG and the proposed mSAA shift the revenue–surplus operating point diagonally toward a more favorable region \emph{(best performance point: approximately surplus=0.8 for revenue=0.7)}, improving both metrics simultaneously and broadening the set of attainable outcomes; this expansion enables even low-priority or weak-channel devices to secure STFS segments with minimal impact on gateway revenue due to more efficient price/assignment coupling}.

\begin{figure}[t!]
\centering
\includegraphics[width=0.8\columnwidth, trim={0.0mm 2.0mm 9mm 12mm},clip]{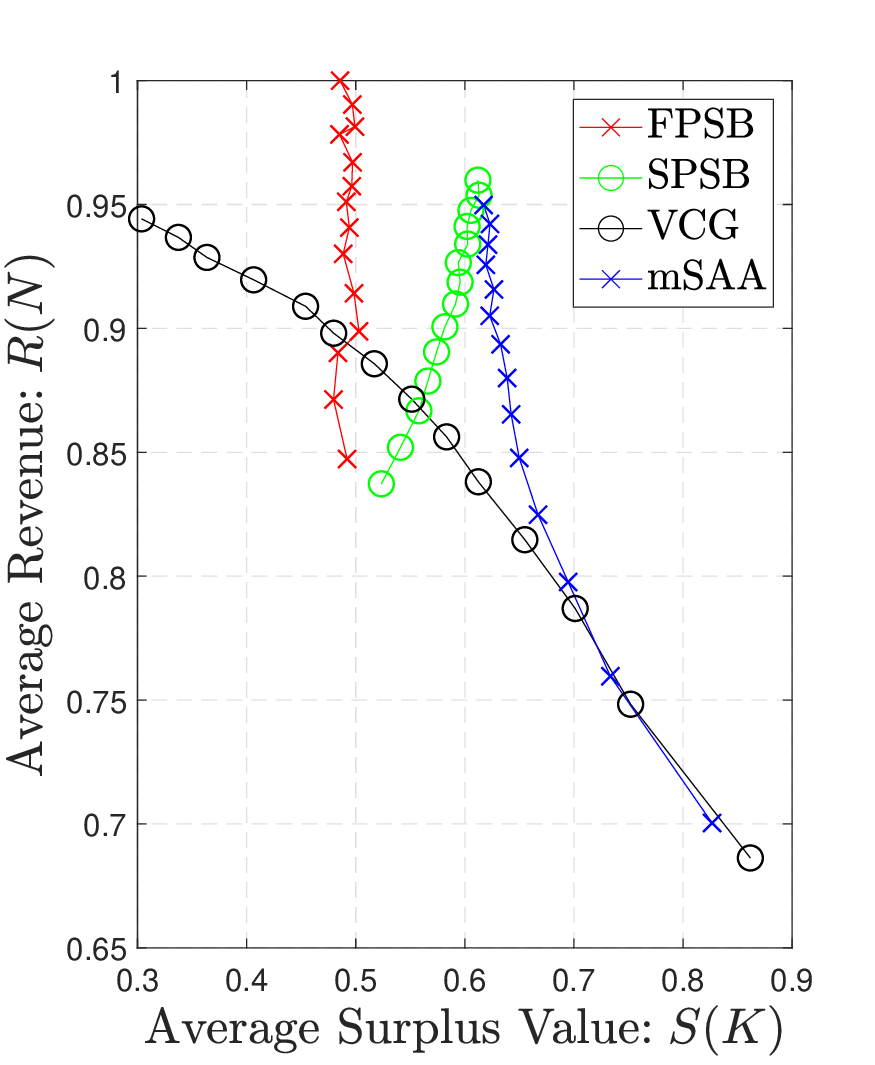}\\
\caption{Relation between normalized average revenue and the normalized average surplus ($\bar{R}$ vs $\bar{S}$) with the varying number of IoT nodes $K$ in the network for $\mathcal{V}=\alpha \mathcal{V}^{H} + \beta \mathcal{V}^{G}$ and $\alpha=1$, $\beta=1$, $a=0$, $b=1$, $\sigma^{2}=1$, $N=10$ over underbidding ($\zeta = 4\%$) options.}
\label{Fig: R_S_zeta_4} 
\end{figure}
Under strategic behavior, \figurename~\ref{Fig: R_S_zeta_4} (risk stage) shows a rightward shift in the revenue–surplus cloud, indicating simultaneous increases in both gateway revenue and node surplus relative to the risk-free case. FPSB exhibits a mild improvement but tends to pin node surplus at an almost fixed level across the revenue range, limiting attainable gains. SPSB demonstrates a pronounced boost in both surplus and revenue emanating from dense (large-$K$) regions, albeit with reduced flexibility across operating points. VCG performs poorly under misreporting, with scattered outcomes reflecting lower surplus and sometimes higher revenue due to added social-cost (Clarke) penalties in underbidding scenarios; its sensitivity to deviation from truthfulness degrades welfare. \textbf{The proposed mSAA dominates across regimes \emph{achieve approximately double surplus compared to VCG for revenue=0.95}: its outcomes envelop the high-performing SPSB region and extend into broader surplus-maximizing zones, evidencing better adaptation to the dynamics of densely deployed IoT settings where bidder strategies and effective channel conditions co-evolve with prices.}

The revenue–surplus frontier represents how the system allocates scarce orthogonal resources among links with heterogeneous instantaneous SNR and valuation profiles. Moving along the frontier corresponds to trading off where power is spent in the network: points with higher revenue typically coincide with allocations that push more traffic onto stronger links (or induce higher power on winners), improving received signal strength at the gateway and reducing collisions; this extracts more payment (revenue) but leaves less surplus for devices. Mechanisms like mSAA and (under truthful play) VCG realign prices with link quality and valuation structure so that STFS segments are preferentially awarded to energetically efficient links, which lifts both the gateway’s collected revenue and the network’s productive surplus—hence the diagonal improvement. When misreporting is allowed, strategic devices can tilt allocations toward otherwise weaker links; physically, this raises the transmit energy required to meet rate targets and perturbs interference patterns, which explains VCG’s welfare collapse under underbidding. In contrast, mSAA’s monotone price updates act like iterative Lagrange multipliers that dampen energetically inefficient matches over iterations, re-concentrating assignments on higher-SNR (or higher-valuation) links and thereby expanding the practically achievable revenue–surplus region even in risk-prone environments.

\subsection{Resource Allocation metrics}
\begin{figure*}[t!]
\centering
$\begin{array}{ccc}
\includegraphics[width=0.316\textwidth, trim={0.0mm 2.0mm 9mm 11mm},clip]{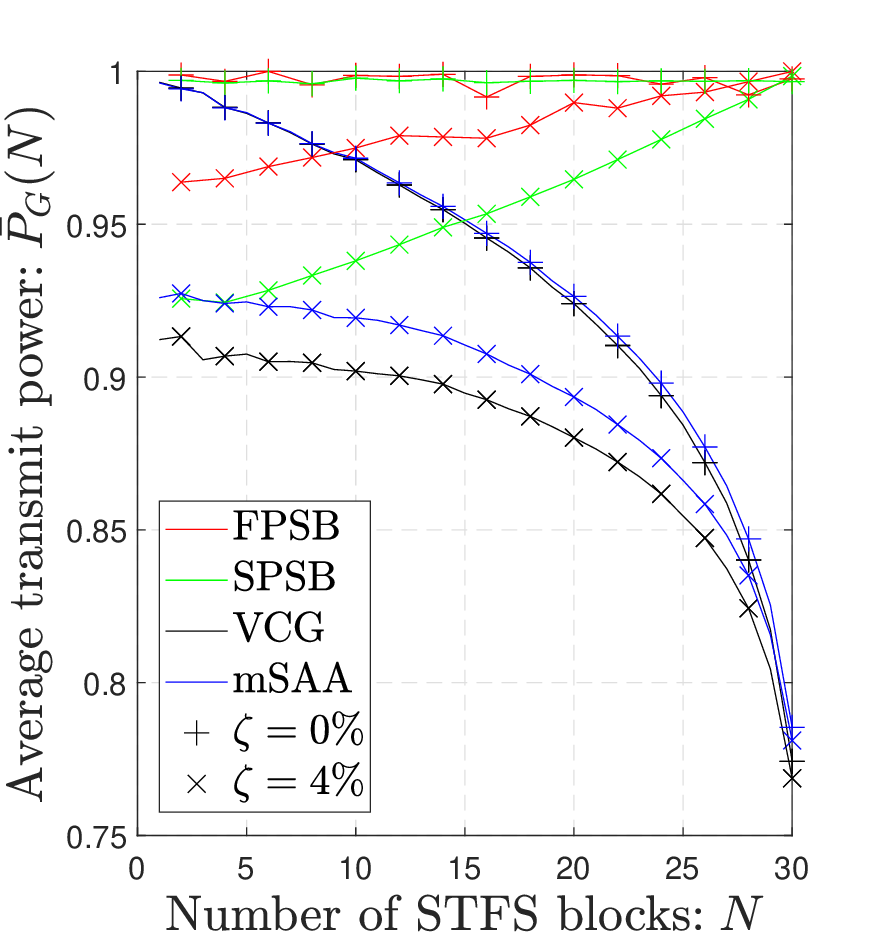} &
\includegraphics[width=0.312\textwidth, trim={1.6mm 2.0mm 9mm 11mm},clip]{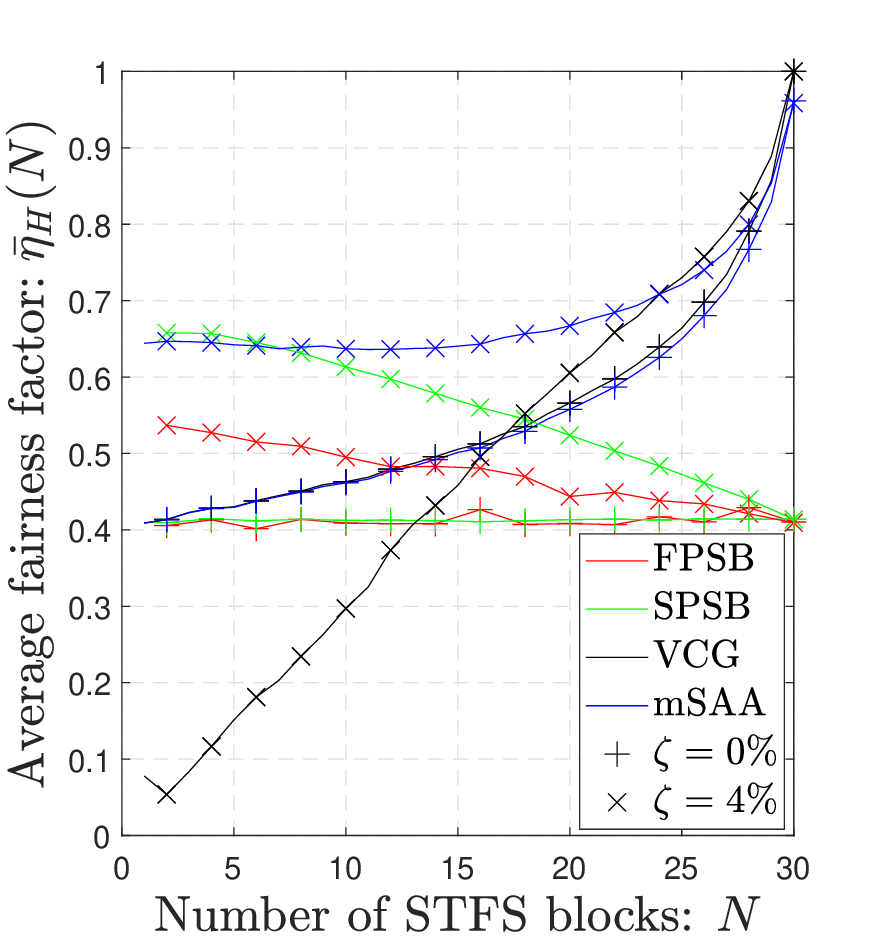} &
\includegraphics[width=0.32\textwidth, trim={0.0mm 2.9mm 9mm 11mm},clip]{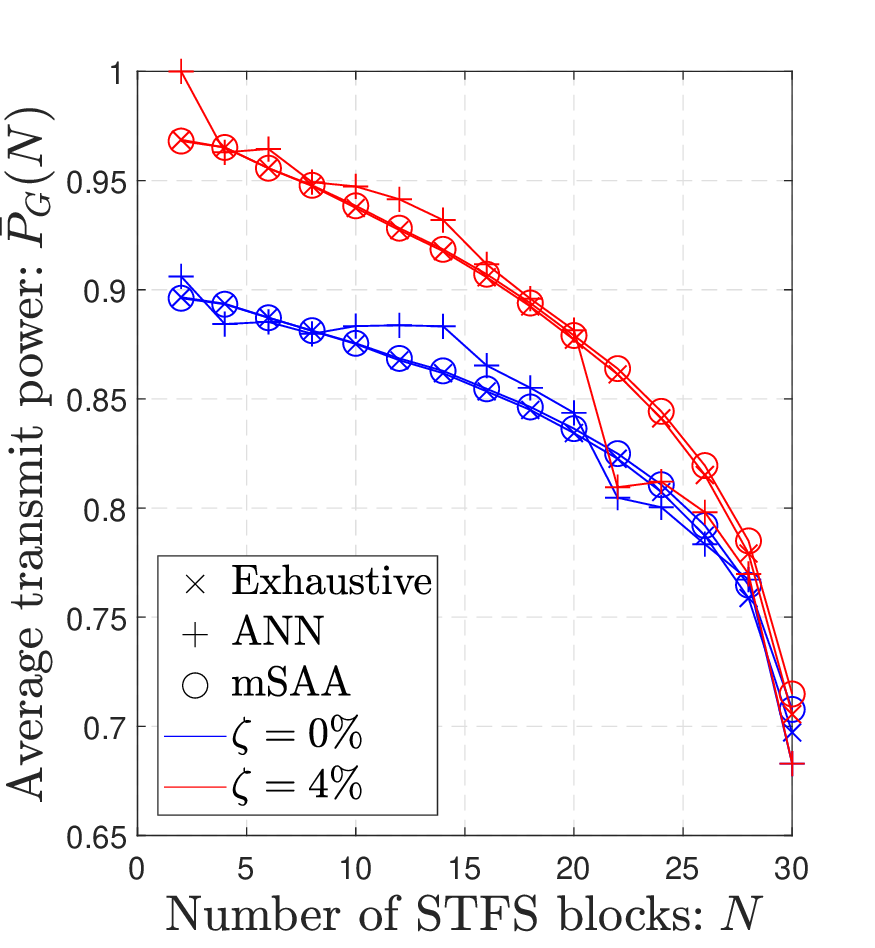}
\\
\mbox{({\textit{a}})} & \mbox{({\textit{b}})} & \mbox{({\textit{c}})}\\
\end{array}$ 
\caption{({\textit{a}}) Normalized average transmit power gain ($\bar{P}_{G}$ vs $N$), ({\textit{b}}) Normalized average resource allocation fairness factor ($\bar{\eta}_{H}$ vs $N$), and ({\textit{c}}) Performance comparison in Normalized average transmit power gain ($\bar{P}_{G}$ vs $N$), aginst to the scalability of STFS slots at the gateway for $K=31$ IoT devices in the network under $\mathcal{V}=\alpha \mathcal{V}^{H} + \beta \mathcal{V}^{G}$ and $\alpha=1$, $\beta=1$, $a=0$, $b=1$, $\sigma^{2}=1$, for ({\textit{a}}), ({\textit{b}}) over risk-free bidding ($\zeta = 0\%$) and underbidding ($\zeta = 4\%$) scenarios then, for ({\textit{c}}) over risk-free bidding ($\zeta = 0\%$) and overbidding ($\zeta = 4\%$) scenarios.}
\label{Fig: pG_N_fH_BM_pG_N}
\vspace{-4mm}
\end{figure*}
\figurename~\ref{Fig: pG_N_fH_BM_pG_N}$(a)$ shows that as STFS capacity increases, node competition decreases, lowering average power consumption. While FPSB and SPSB maintain high transmission power regardless of competition, VCG and mSAA significantly reduce power usage, especially in less competitive regions (\( K/2 < N \leq K \)). However, underbidding (\( \zeta = 4\% \)) disrupts RET, affecting both classical (FPSB, SPSB) and extended (VCG, mSAA) mechanisms. Despite this, the allocation indicator \( c_{\mathcal{K}}^{\mathcal{N}} \) still favors node-side benefits, lowering transmission energy by awarding STFS blocks to nodes with strong channels. Risk-taking in dense networks or scarce resource conditions (\( N < K \)) may violate conventional water-filling behavior, but as \( N \lesssim K \), power usage trends return to stable, risk-free levels across all auction types.


\figurename~\ref{Fig: pG_N_fH_BM_pG_N}$(b)$ shows the normalized average resource allocation fairness factor for IoT devices relative to the available STFS block capacity at the gateway. Higher-priority nodes with stronger uplink gains tend to access the spectrum more frequently. Classical FPSB and SPSB auctions fail to improve fairness based on individual priorities under standard bidding. In contrast, VCG and mSAA significantly enhance fairness by leveraging diverse node hypotheses, especially as more heterogeneous resource slots become available. When nodes underbid (\( \zeta = 4\% \)) to save transmit power, fairness improves across all mechanisms—most notably in highly competitive regions with limited STFS blocks. Traditional FPSB and SPSB gain short-term equity among risky nodes, but this fades in risk-free scenarios. Although VCG slightly outperforms mSAA in reducing power use (\figurename~\ref{Fig: pG_N_fH_BM_pG_N}$(a)$), it falls behind in fairness (\figurename~\ref{Fig: pG_N_fH_BM_pG_N}$(b)$) due to social cost penalties on low-reputation nodes, disrupting equilibrium. Ultimately, mSAA proves more robust, enabling fairer, more frequent STFS assignments to high-priority nodes. It effectively supports dense and diverse edge environments, even under asymmetric CSI conditions.

\subsection{Performance Comparison}
Finally, we evaluate the transmit-power efficiency of the proposed mSAA against two baselines: (i) a global reference obtained by exhaustive search, and (ii) a four-layer dense Artificial Neural Networks (ANN) model trained on those global-reference labels \cite{our_GLOBECOM_paper}. The brute-force solver treats allocation as a *perfect-information* auction and enumerates all \(\binom{K}{N}\) candidate assignments (lowest-density case \(K=N\)), with computational burden reported as \(\mathcal{O}(N!\cdot N)\). The ANN baseline is then supervised on these optimal labels to learn a direct mapping from features (channel/hypothesis vectors) to resource assignments. \figurename~\ref{Fig: pG_N_fH_BM_pG_N}(c) reports normalized power-efficiency for both risk-free bidding and an overbidding stress test with misreporting level \(\zeta=4\%\). In the latter, valuations are inflated according to, $v_k' \in \mathcal{V}_{\zeta}\; \leftarrow\; \bigl\{\, v \in \mathcal{V}\,\bigm|\, v + \max(v)\cdot \zeta\% \bigr\}$. Overbidding enables low-priority or weak-channel nodes to secure STFS slots more frequently; however, this comes at the cost of higher transmit energy, as indicated by the red traces in the figure. Across both bidding regimes, the proposed mSAA tracks the exhaustive optimum closely, while the ANN exhibits larger deviations from the global reference—highlighting that a learned, one-shot predictor trained on fixed labels can struggle to generalize under strategic behavior shifts, whereas the iterative, price-responsive mSAA maintains near-optimal power efficiency.

The exhaustive-search baseline functions as an oracle: it always selects the assignment that minimizes transmit power (or maximizes the stipulated efficiency metric) given the instantaneous channel/hypothesis realizations. The proposed mSAA approaches this oracle by continuously adjusting prices and allowing devices to re-evaluate participation at each iteration; economically, the price vector $\mathbf{q}$ acts as a Lagrange multiplier that nudges the market toward allocations where devices with inherently better channel conditions (or higher true valuations) win orthogonal STFS segments. Physically, this alignment steers traffic onto links with higher effective SNR and better dispersion compensation, thereby lowering the aggregate transmit energy needed to meet throughput constraints. When nodes overbid $\zeta>0$, devices with weaker links artificially inflate their valuations to obtain resources. This shifts allocation probability toward energetically disadvantageous links: more power is required to sustain target rates over poorer channels and to counteract residual interference or dispersion mismatch. The observed red curves therefore move upward (higher energy), capturing the physical cost of assigning scarce orthogonal segments to channels with lower gain.

The ANN baseline departs from the oracle under both truthful and strategic regimes because it performs a single, fixed mapping learned from historical labels. Without an internal price-adjustment loop, the ANN cannot react to distributional shifts induced by misreporting $\zeta>0$ or to subtle coupling between users’ choices across iterations. In physical terms, it can assign resources to suboptimal links more often, which manifests as higher required transmit power. By contrast, mSAA’s iterative price monotonicity filters out energetically inefficient matches over time, converging to allocations that are closer to the oracle and thus more power-efficient in practice.


Although the accuracy of ANNs can be enhanced with richer training data, they often struggle to handle multi-dimensional CSI and node hypotheses, and they lack adaptability to dynamic conditions such as underbidding or overbidding. In other words, the performance of ANN-based resource allocation heavily depends on the quality and diversity of the training dataset. However, in dense and heterogeneous IoT networks, there exist numerous classes of incomplete parameters that must be considered during training. Furthermore, the weights in a simplified ANN model are typically optimized for specific input data vectors tailored to a particular use case, which can result in outliers when unexpected bidding behaviors occur among users. In contrast, the proposed mSAA algorithm achieves a near-optimal resource allocation performance, closely approximating the results of exhaustive search methods. This ensures high performance even in highly competitive scenarios with limited STFS resources. Additionally, the mSAA-based allocation remains efficient, scalable, and robust under varying auction behaviors, including overbidding.
\subsection{Extreme Misreporting} 
\begin{figure*}[t!]
\centering
$\begin{array}{cc}
\includegraphics[width=0.35\textwidth, trim={0.0mm 2.0mm 11mm 12mm},clip]{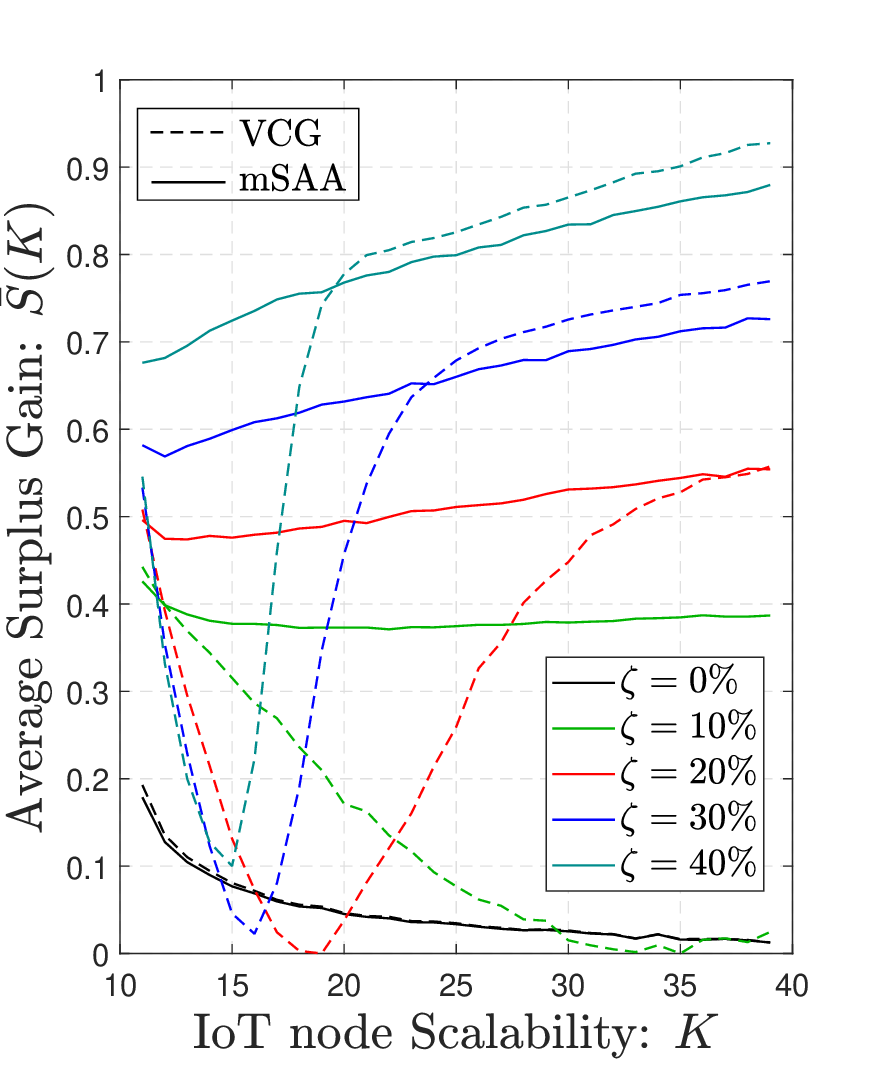} &
\includegraphics[width=0.35\textwidth, trim={0.0mm 2.0mm 11mm 12mm},clip]{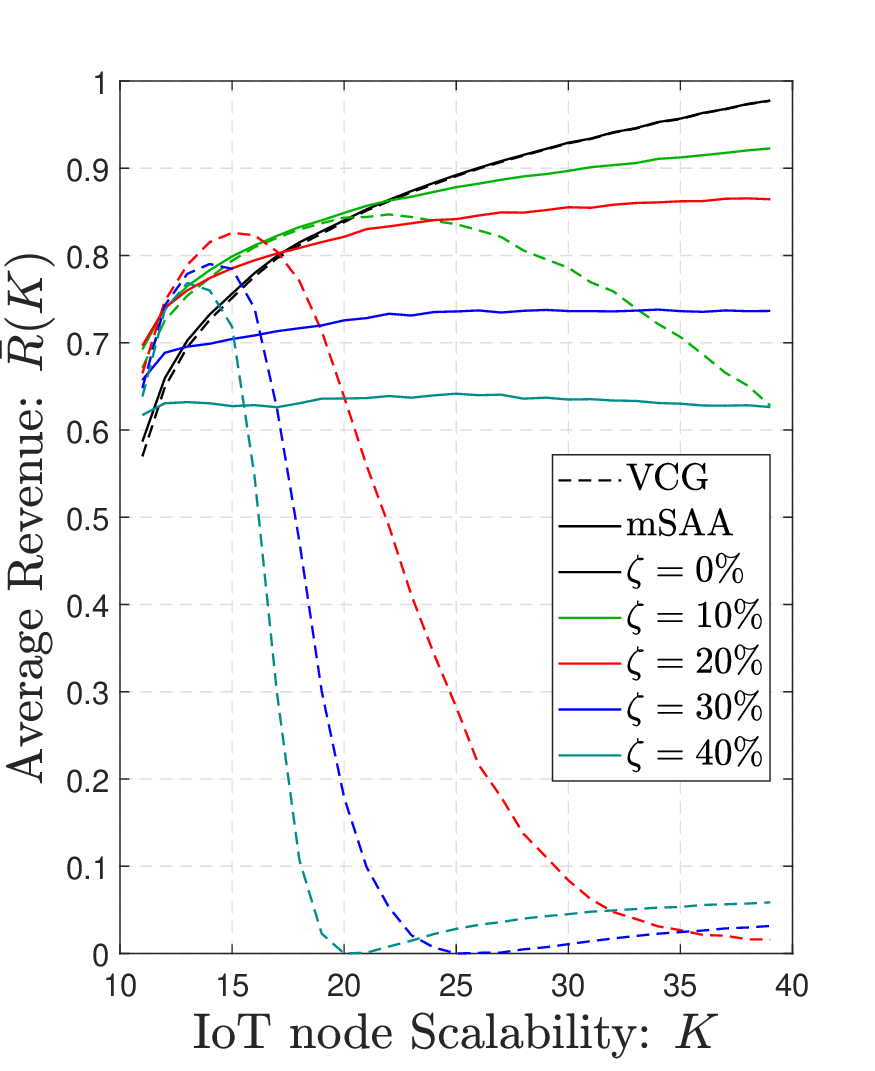}
\\
\mbox{({\textit{a}})} & \mbox{({\textit{b}})}\\
\end{array}$ 
\caption{({\textit{a}}) Normalized average surplus ($\bar{S}(K)$ vs $K$), and ({\textit{b}}) Normalized average revenue ($\bar{R}(K)$ vs $K$), aginst to the scalability of IoT devices in the network for $\mathcal{V}=\alpha \mathcal{V}^{H} + \beta \mathcal{V}^{G}$ and $\alpha=1$, $\beta=1$, $a=0$, $b=1$, $\sigma^{2}=1$, $N=10$ over extreme underbidding stages $\zeta~\in~\{ 0\%, 10\%, 20\%, 30\%, 40\% \}$.
}
\label{Fig: Surplus_Revenue_K_zeta}
\vspace{-4mm}
\end{figure*}
\figurename~\ref{Fig: Surplus_Revenue_K_zeta}(a) reports the average surplus gain versus network scalability for VCG and the proposed mSAA under \emph{extreme underbidding} levels. Under the standard (truthful) bidding strategy ($\zeta=0\%$), VCG and mSAA yield similar surplus magnitudes, and both curves decrease as $K$ grows (denser competition). When bidders deviate strategically with larger $\zeta$, IoT nodes can increase their \emph{self-surplus} by submitting reduced valuations, i.e., $v_k' \in \mathcal{V}_\zeta \leftarrow \{\, v \in \mathcal{V} \mid v - \max(v)\cdot \zeta\% \,\}$, thereby lowering payments conditional on still winning. In this regime, the mSAA architecture exhibits an \emph{uplift} of the surplus curves with positive rates even at higher $K$, indicating resilience of the allocation/pricing dynamics to misreporting. In stark contrast, surplus under VCG drops sharply for extensive underbidding because the Groves/Clarke mechanism imposes higher social-cost adjustments (taxes) on allocations that become misaligned with true externalities, leading to a pronounced welfare loss. Beyond a threshold (empirically around $\zeta>10\%$), the measured surplus increases markedly toward the best-performing mSAA levels; here, the strength of misreporting surpasses the corrective effect of the social-cost term in~\eqref{Eq: Grove_mechanism_2}, producing \emph{turning points} whose locations vary with $\zeta$. In some configurations, the surplus attained by underbidding nodes slightly surpasses the nominal maximum achievable under the baseline mSAA (truthful) configuration, reflecting that the \emph{surplus metric} includes a payment component that is directly reduced by underbidding.

\figurename~\ref{Fig: Surplus_Revenue_K_zeta}(b) shows the corresponding gateway revenue. Under truthful bidding ($\zeta=0\%$), VCG and mSAA deliver comparable revenue that increases with $K$. As $\zeta$ grows beyond $10\%$, the revenue growth rate of mSAA tapers because nodes submit lower valuations, which compresses price trajectories and payments even as competition persists. \textbf{The conventional VCG mechanism, however, fails to preserve revenue under severe underbidding and exhibits a substantial decline, consistent with the surplus degradation observed in~\figurename~\ref{Fig: Surplus_Revenue_K_zeta}(a): \emph{Surplus under extreme underbidding e.g., for $\zeta=20\%$, the average surplus gain of mSAA ($\approx0.5$) is approximately doubled compared to VCG ($\approx0.25$) and in parallel, the average revenue gain of mSAA ($\approx0.85$) is exceeding twofold compared to VCG ($\approx0.3$) at node scalability K=25}.}
Taken together, \figurename~\ref{Fig: Surplus_Revenue_K_zeta} demonstrates that VCG’s overall performance degrades extensively under strategic misreporting—yielding lower IoT surplus and lower gateway revenue—whereas mSAA remains robust and sustains favorable performance across both metrics.

Under extreme underbidding, devices intentionally report valuations below their true link utility. Physically, this tends to reassign some orthogonal STFS resources toward links that are not the strongest instantaneous-SNR options and to lower the clearing prices paid by winners. In VCG, the Clarke pivot tax is calibrated for truthful reports: when reports are distorted, the social-cost correction misprices externalities, so assignments drift away from energetically efficient links, increasing required transmit power and reducing realized welfare and gateway revenue. By contrast, mSAA’s monotone price ascent and active-loser filtering act like iterative Lagrange multipliers: even with depressed bids, prices adjust over iterations to favor allocations that are \emph{feasible} and \emph{energetically efficient}, gradually steering STFS segments back toward higher-SNR or higher-effective-valuation links. The observed uplift of the \emph{surplus} metric for large $\zeta$ under mSAA arises because surplus equals valuation minus payment; underbidding primarily suppresses the payment term while mSAA’s allocation rule still captures much of the physical channel advantage. Consistently, \figurename~\ref{Fig: Surplus_Revenue_K_zeta}(b): \emph{Revenue under malicious underbidding} 
shows that this uplift is accompanied by a slower revenue increase (or decline for VCG), confirming the energy–economics trade: truthful, price-aligned allocations concentrate power on favorable channels and support revenue, whereas severe misreporting diverts resources toward weaker links, increases the physical energy required to meet rate targets, depresses revenue, and—absent adaptive pricing—erodes system welfare.


\section{Conclusion} \label{Sec: 7_conclusion}
In this paper, our auction-based proposed mechanism (mSAA) facilitates a federated STFS-based resource allocation optimization task with incomplete CSI and individual hypotheses among IoT devices. This scalable approach is applicable to diverse IoT tiers in the resource allocation framework while capturing specific entity interactions to distribute the computational complexity in a surrounding with competitive demands. Additionally, optimal dispersion vectors involve prior compensation of environmental uncertainty for IoT firings to reduce interference at the gateway. The proposed framework outperforms both conventional auction models and the baseline ANN model by allocating minimal transmit power levels to IoT entities, while preserving system integrity even under asymmetric user bidding behaviors. 
In the future, we plan to significantly sharpen the prediction capabilities of bidding players within the mSAA architecture. We aim to achieve this by leveraging advanced graph signal processing techniques to model and analyze the internal self-behaviors of IoT devices, which should lead to more precise forecasting of bidding patterns and resource demands. We will also delve into the numerical derivation of optimal bidding strategy vectors for the mSAA mechanism. This will involve utilizing order statistics of bidding vectors and incorporating VCG price rules alongside RET within standard risk-free game models. This will establish a more concrete and computationally verifiable framework for optimal resource allocation in competitive IoT environments. Finally, we will extend our focus to the broader implications of mSAA for mobile and 6G networks. We will investigate how the decentralized nature and inherent robustness of mSAA can directly address critical challenges in next-generation wireless communication, particularly in areas like efficient spectrum management in licensed shared access (LSA) systems and ensuring fair resource allocation for diverse user demands. Our goal is to solidify mSAA's role as a foundational mechanism for economically inspired, user-centric resource allocation in the ultra-dense and dynamic landscapes anticipated for future wireless networks.

\bibliographystyle{IEEEtran}
\bibliography{IEEEabrv,ref}
\end{document}